\documentclass[preprint]{JHEP}

\usepackage{epsfig}

\def\be{\begin{eqnarray}}		
\def\ee{\end{eqnarray}}
\def\0{\nonumber}
\def\ba{\begin{array}}			
\def\ea{\end{array}}
\def\bv{\left(\mkern-9mu\ba{c}}		
\def\ev{\ea\mkern-9mu\right)}
\def\bs{\begin{tabular}{|c|}\hline$}	
\def\es{$\\ \hline\end{tabular}}
\def\bma{\left(\begin{tabular}}		
\def\ema{\end{tabular}\right)}
\def\bc{\begin{center}}
\def\ec{\end{center}}

\def\dtre{{\rm d}^3}
\def\di{{\rm d}}
\def\<{\left<}
\def\>{\right>}
\def\e{{\rm e}}
\def\tg{{\rm tan}}

\def\tr{{\rm tr}}
\def\Tr{{\rm Tr}}
\def\Det{{\rm Det}}

\def\d{{\partial}}
\newcommand{\dd}[1]{{\partial \over \partial #1}}
\def\.{\dot}

\def\al{{\alpha}}
\def\lm{{\lambda}}

\def\id{{\bf 1}}
\def\p{{\bf p}}	
\def\q{{\bf q}}	
\def\m{{\bf m}}

\def\a{{\mbox{\boldmath$\alpha$}}}
\def\ch{{\mbox{\boldmath$\chi$}}}
\def\ph{{\mbox{\boldmath$\phi$}}}
\def\vph{{\mbox{\boldmath$\varphi$}}}
\def\lamb{{\mbox{\boldmath$\lambda$}}}

\def\gmunu{g_{\mu\nu}}
\def\Amu{A_\mu}

\def\amu{a_\mu}
\def\anu{a_\nu}
\def\Dmu{D_\mu}
\def\Dnu{D_\nu}

\def\h{\hbox{$\cal H$}}

\def\T{{\bf T}}
\def\A{{\bf A}}

\def\R{{\rm R}}
\def\Z{{\rm Z}}

\def\D{{\cal D}}
\def\S{{\cal S}}

\def\cg{{\cal G}}
\def\cM{{\cal M}}
\def\cN{{\cal N}}

\newlength{\extraspace}
\newlength{\extraspaces}
\newcounter{dummy}
\newcommand{\bei}{
	 \addtocounter{equation}{1}
	 \setcounter{dummy}{\value{equation}}
	 \setcounter{equation}{0}
	 \renewcommand{\theequation}{\thesection.\arabic{dummy}\alph{equation}}
	 \begin{eqnarray}
	 \addtolength{\abovedisplayskip}{\extraspaces}
	 \addtolength{\belowdisplayskip}{\extraspaces}
	 \addtolength{\abovedisplayshortskip}{\extraspace}
	 \addtolength{\belowdisplayshortskip}{\extraspace}}
\newcommand{\eei}{
	 \end{eqnarray}
	 \setcounter{equation}{\value{dummy}}
	 \renewcommand{\theequation}{\thesection.\arabic{equation}}}

\newcommand{\bmp}[1]{\noindent
                     \begin{minipage}[b]{#1}
                     \setlength{\parindent}{.5cm}\vspace*{5pt}}
\newcommand{\emp}[1]{\end{minipage}\raisebox{10pt}{\hspace*{10pt}
                     \includegraphics{#1}}\hspace*{\fill}\linebreak[3]}

\title{Three dimensional large $N$ monopole gas}

\author{Fabrizio Nesti \\ SISSA-ISAS, Trieste, Italy. 
			  E-mail: \email{nesti@sissa.it}}


\preprint{SISSA/ISAS ref: 143/96/EP\\\hepth9610012}

\abstract{
We study here the large $N$ limit in the presence of magnetic monopoles
in the Yang-Mills/Higgs model in three dimensions. The physics in the
limit depends strongly on the distribution of eigenvalues of the
Higgs field in the vacuum, and we propose a particular, nondegenerate
configuration. It minimizes the free energy at the moment of symmetry
breaking. Given this, the magnetic monopoles show a wide hierarchy of
masses, and some are vanishing as $1/N$. The dilute gas picture, then,
provides an interesting framework for the large $N$ analysis. }

\begin{document}

\section{Introduction}

The subject of this paper is the analysis of some nonperturbative
features of the Yang-Mills-Higgs (YMH) model in three dimensions for gauge
group $SU(N)$ and some implications for the large $N$ limit. 

\medskip
Our YMH model is the direct generalization to $SU(N)$ of the 
Georgi-Glashow model \cite{GeoGla}. This was one of the first models 
providing a spontaneous breaking of a gauge symmetry.

The vacuum expectation value of the Higgs field provides a length scale 
and the size for magnetic monopoles, providing their stability. 

It was however argued in the known 't Hooft paper on oblique confinement 
\cite{tHooft} that such configurations should exists also in 
pure Yang-Mills theory of strong interactions, and play a major role 
in the mechanism of quark confinement with their condensation. 

\medskip
Already the Georgi Glashow model in two dimensions shows analogous
configurations, {\it vortices} \cite{NieOle}, which account for the 
BCS theory of superconductivity \cite{Weinberg}.

It was an idea from the early ages of QCD \cite{conftHo,confMan,confPar} 
that quark confinement could be explained with a picture analogous to 
that of type II superconductors but with the role of electric and magnetic
charges interchanged (and for late issues like abelian dominance in confinement
see \cite{DGL-conf, DGL-conf2}).

For nonabelian groups $SU(N)$ the vortex solutions are not stable, 
mainly due to the triviality of the fundamental group, and this 
complicates the matter in that it requires the breaking to some 
subgroup with abelian factors. 

\medskip
This is exactly realized in the YMH model, and the infrared behavior
of gauge theories with compact gauge group \cite{Poly75}, was explicitly
shown to hold in the context of the Coulomb gas approach to the ensemble 
of widely separated monopole solutions \cite{Poly77}. 

This treatment is based on the fact that widely separated monopoles 
interact via a Coulomb interaction, despite the exact form of the 
energy is not known explicitly for generic parameters 
and also the space of moduli of the generic solutions 
has a nontrivial geometric structure \cite{AtiyahHitchin}. 

After these early years, lattice simulations have been largely employed to
test the occurrence and condensation of monopoles in correspondence to the
deconfinement transition, in the pure Yang-Mills theory \cite{digiacMCTV}.

Because of the absence of the Higgs field, monopoles are of no fixed size,
and thus some mechanism of generation of mass is necessary if monopoles 
are to be relevant for the confinement mechanism. This is indeed found on
the lattice and there are arguments in the three dimensional continuum
theory \cite{DasWadia94}.

Let's recall also that the {\it static} four dimensional theory is 
equivalent to the YMH model with the limit of no potential. 
This is one of the main reasons to study the YMH model. 

\bigskip
Many interesting features are in the domain of the large $N$ expansion, 
mainly:
a) the classification of Feynman graphs according to the surfaces where
they can be embedded gives also the order in $1/N$, actually $1/N^2$, 
of the graph and this allows the contact with the string interpretation 
of gauge theories on one hand, and on the other with the random surface 
interpretation of zero dimensional matrix models; 
b) for suitable operators \cite{largeN2} the correlation functions 
``factorize'', that is, are given by the disconnected part at 
leading order; c) finally $N$ is not renormalized, in that 
it is a dimensionless external fixed parameter. 

Of course a special mention is due to the pure $N=\infty$ case,
where one implicitly lets $N\to\infty$ {\it before} removing any
cutoff. This can bring the theory to have a different
phase structure. There is a specific example in one dimension 
that is the Kazakov-Migdal phase transition. 

The second property above, factorization, shows that for the operators
which follow it, the large $N$ limit is a kind of semiclassical
limit. Their fluctuations are suppressed, and $1/N$ plays the role 
of $\hbar$ for them.

\medskip
One of the recent interesting phenomena, is the emergence of a new 
dynamic in the $N=\infty$ limit of some theories. 
These include affine Toda models, principal chiral fields in one and 
two dimensions, two dimensional QCD, not to mention the matrix models.

In all these theories the effect of taking $N$ large is to generate
an infinity of states which coalesce to form collective excitations 
of some other, higher dimensional, theory. 

The emergence of a new dimension is the remnant of the matrix index of
the diagonal fields. In these theories angular variables play an
important role but are integrated, more or less explicitly, to leave
an effective interaction for the eigenvalues.

An example is the theory of the principal chiral field one the
line\cite{kazwin}, which is equivalent, for some boundary conditions, 
to two dimensional QCD with its interpretation in terms of two
dimensional string. 

\medskip
We come thus to our case of Coulomb gas of magnetic monopoles. 
It was constructed as a sum on the classical dilute configurations of 
monopoles, weighted with the determinant of gaussian fluctuations 
around them. 

With the Coulomb gas there is, via a Sine-Gordon transform, a dual 
representation and the possibility to achieve an estimate about the 
{\it string tension} for Wilson loops of large area. 

The string tension is related to the pseudo-mass of the monopoles by an
exponential relation, which shows, like in the dual Landau-Ginzburg
theory, that it is related to the density of monopoles.

\medskip
For $SU(N)$ we have $N(N-1)$ species of minimal monopoles, with magnetic 
charges in different couples of $U(1)$ sectors, and the coulomb gas can be
generalized to this case \cite{DasWadia}, as well as the Sine-Gordon transform.

One has however to take into account the different species of monopoles 
that populate the model, and this was not done in \cite{DasWadia}.
Accordingly one has to consider the determinant of quantum fluctuations 
around the different monopole backgrounds. 

Up to this point the analysis is valid for any $N$. 

\medskip
A possible new behavior comes from the large $N$ limit,
because there necessarily appears a distribution of masses which 
are present in the theory. 

These masses can be of order, a priori, in the interval from $0$
to $N$, and the physics at large $N$ of course has to be very different 
from case to case.

In this model the monopole pseudo-masses and the mass of the gauge bosons
are governed by the differences of eigenvalues of the Higgs field at infinity,
$\phi^\infty$, so that all the model depends on its distribution of 
eigenvalues {\it in the vacuum}.

\medskip
In the standard picture of symmetry breaking $\phi^\infty$ can not be changed
by any fluctuation, once the universe has formed. One simply fixes it.
The modulus of $\phi^\infty$ gets renormalized and possibly shifted 
as with the arguments of the effective Higgs potential, but there is no
indication on its direction in the Cartan space. 


\medskip
However in the course of our analysis we are lead to use the
{\it unitary gauge}, because the physical degrees of freedom are
explicit and the Higgs field is diagonal, and a curious phenomenon arises.

The unitary gauge is somewhat singular, because its Faddeev Popov 
determinant is not defined in the continuum. This turns out to be the 
product at each point of the Vandermonde determinant constructed 
with the eigenvalues of $\phi^\infty$. 

This factor in the functional integral seems to provide measure zero
for all configurations where some eigenvalues coincide, and to give a sort
of repulsion of eigenvalues. 

This immediately faces with the problem that monopole
configurations, which have necessarily points where the eigenvalues
coincide, have all zero weight. 

\medskip
Fortunately as the analysis is carried on, and still thinking that the
theory {\it is} renormalizable, we can show that once the quantum
fluctuations of the massive gauge fields are taken into account, 
the Van-der-Monde ultralocal determinant is canceled almost completely. 
What remains is just the Van-der-Monde determinant of the eigenvalues
of $\phi^\infty$!

\medskip 
Still this term does not authorize us to think to a `quantum lifting' 
of the degeneracy in Cartan directions, because again $\phi^\infty$ is 
fixed at the ``beginning of the universe''.

However we think that at the epoch of its formation, the system
{\it is} sensible to this term, and thus chooses the distribution
of eigenvalues which maximizes it. 

We have found this distribution, which determines back the
distribution of masses of gauge bosons and monopoles in the system.

It shows a peculiar shape, and should bring peculiar consequences in
the properties of the system.

\medskip
It is worth noting that the same distribution of eigenvalues of 'Higgs 
field' and gauge boson masses, is found \cite{DouShe} in the recent 
nonperturbative solution of $SU(N)$ supersymmetric Yang-Mills theory 
in four dimensions (the $\cN=1$ case), once the complete confinement is
required. 
It is striking that the same distribution appears in the large $N$, 
and as its very origin is physically not known in that case.

\medskip
Coming back to our problem of the monopole gas, we have now a
distribution to analyze the system and we could proceed. 

Unfortunately the very difficulty to the completion of the program 
is still there, because the semiclassical sum needs some estimate with $N$ 
of the determinant in the external field of a monopole. 

This problem is still unsolved despite many efforts
\cite{Oleszcz, det1, det2, zarembo} and is nontrivial. 
An approach with Eguchi Kawai reduction which could provide at least the large $N$ behavior is being studied.

\bigskip
Let's summarize our path in this work.

The first part of this work deals with some new large $N$ ideas for 
the YMH model in three dimensions. Needless to say there is a large 
historical and scientific background and it is of course difficult 
to say something really new on these subjects. Nevertheless some
latest ideas on matrix models, For example, to our knoweledge, 
the study of Coulomb gases at large number of species is not investigated.

\medskip
After one finds a reliable method to estimate the functional determinant, 
it will be possible to draw definite conclusions on the large $N$
monopole gas, which seems to promise interesting features, due 
to the competition of factors in the functional integral 
which takes place in the large $N$ limit.

\bigskip

\section{Preliminaries}\label{ym3}

We start by considering the Yang-Mills/Higgs$_3$ model built on the
gauge group $SU(N)$, so it is a $3$-dimensional euclidean theory of
a gauge field $A_\mu=A_\mu^aT_a=\A_\mu\cdot\T$ and a matter field
$\phi=\phi^aT_a=\ph\cdot\T$ both living in the adjoint representation.

Both are arranged in $N\times N$ matrix fields living in the algebra of
$SU(N)$. In all the following we will always denote with plain symbols
such fields, like $A_\mu$, $\phi$, and with bold symbols
the vector of their components along the algebra generators, like
$\A_\mu$, $\ph$ above. Whenever the field will be diagonalized, its
vector will have only the Cartan components.

To fix the notation we take the normalization such that
$\tr(T_aT_b)={1\over2}\delta_{ab}$ in the fundamental representation.
We will occasionally mention also the
theory with $T_a$ in the adjoint representation, in next chapter, because
they give rise to very different scenario of monopoles. 

The configuration space $\Gamma$ is the space of functions from $\R^3$ to
the couple $(A,\phi)$ with finite action $\S$. 
In this configuration space acts a continuous $SU(N)$ gauge group:
	\be A_\mu(x) &\to& \omega^{-1}(x)\d_\mu \omega(x)
			+g \omega^{-1}(x) A_\mu(x)\omega(x)\\
     \phi(x) &\to& \omega^{-1}(x)\phi(x)\omega(x)\ee
which leaves the action invariant.

\bigskip
Because we will focus on the large $N$ expansion,
it is necessary to adapt the parameters of the theory to this limit.

It is the standard remark \cite{HooVel} that the large $N$ limit 
is nontrivial only if the perturbative series remains finite,
and this requires the coupling constants to be suppressed with
powers of $N$. Rescaling the fields one can require all terms to
be of the same order $N^2$.

The action is then:
  \be \S = \int\dtre x \left\{
{N\over2g^2}\tr(G_{\mu\nu}G_{\mu\nu}) +{N\over2}\tr(\D_\mu \phi\D_\mu\phi)+V(\phi)\right\}\ee
with $\D_\mu = \d_\mu + g [A_\mu,\cdot]$, $G_{\mu\nu} = [\D_\mu,\D_\nu] =
			g(\d_\mu A_\nu-\d_\nu A_\mu)+g^2[ A_\mu, A_\nu]$
and $V(\phi)$ a scalar potential.

This YMH theory in three dimensions can be seen as the static version of
the four dimensional minkowsky YMH, or even of self-dual pure YM theory,
at finite temperature, where the $A_0$ component assumes the part of the
Higgs field (and $V=0$). It has thus also direct phenomenological interest.

So, to make contact with the four dimensional theories, one
respects four dimensional renormalizability, and takes the potential to
be a quartic polynomial in the traces\footnote{The choice of a function 
		which is symmetric in the different Cartan directions
		is not the only one \cite{Li} but is the more natural.}
of the Higgs field:
	\be V(\phi) = \lm[\tr(\phi^2)-\mu^2]^2\ee
From $N$ power counting $\mu$ has to be of order $N$.

Requiring the vacuum to have finite action (finite energy in four
dimensions) $V(\phi)$ induces a spontaneous shift of the Higgs vacuum
value from zero to some $\phi^{\infty}$, and the gauge group $G$
breaks down spontaneously to a subgroup $H$ which leaves
$\phi^{\infty}$ invariant. This is called the {\it little
group}\footnote{%
The little group $H$ is always of the form $T'\times G'$, where $T'$
is some abelian group and $G'$ is a simple subgroup.
In any case it always includes a $U(1)$ subgroup, called the 
{\it electromagnetic group} and generated by the Higgs field itself.}.

\medskip
$\Gamma$ is divided in disjoint sets classified by the
winding of the two-sphere at infinity into the coset $G/H$:
elements of $\pi_2(G/H)$ which is isomorphic to $\pi_1(H)$ because 
$\pi_1(G)$ is trivial.

In case the vacuum Higgs field has all different eigenvalues
the little group $H$ is the maximal abelian subgroup $U(1)^{N-1}$, and
the classification has $N-1$ topological quantum numbers: $\Z^{N-1}$.
Moreover thanks to the vanishing of $\pi_2(G)$, this classification
is gauge invariant, because gauge transformations of $\phi_\infty$
are homotopic to the identity.

\medskip
Then, in the spirit of the semiclassical approach, one
considers the minimum of the action in each set $\hat\phi(x)$, and
expands $\phi$ around it. One has a pointwise breaking of the local
gauge group 
$G$ down to the little group which leaves $\hat\phi(x)$ invariant. 
All gauge fields not belonging to $H$ acquire a mass, 
while those in the little group remain massless.  
At points where the Higgs field has all different eigenvalues, 
the breaking is maximal and the little group is $U(1)^{N-1}$. 

As `t Hooft shows \cite{tHooft}, the manifold of points where two 
eigenvalues of $\phi$ coincide has dimension $d-3$, that is, in three
dimensions, consists of isolated points. There the field
configurations have the properties of magnetic monopoles, and 
in the next section we will review also, in the case of $SU(N)$, how the
topological number which classifies the Higgs field represents the magnetic
charge of the configuration under the broken symmetry group $H$.
(modulo equivalence under the Weyl discrete symmetry).

\medskip
The other terms in the action, namely the pure gauge and the Higgs
kinetic terms, pose no obstruction\footnote{This happens in the
gaugeless limit, where, turning off the coupling to the gauge fields,
the Higgs kinetic term forces $\phi$ to have the same direction at
infinity, thus giving winding zero.}, and configurations of arbitrary 
winding can be explicitly constructed \cite{Sch, Arafune}.

\medskip
Before reviewing in the next section the classification of classical
solutions for $SU(N)$, we make a remark that we will need later 
valid for all the configurations of nontrivial winding.

Following $\phi$ smoothly in all the space, one necessarily meets points
where it has at least two coinciding eigenvalues, because otherwise the
winding would have disappeared.

This statement is obviously gauge invariant so it is true even in non
regular gauges like the unitary gauge.

\section{Classical monopoles}

In this section we review the classification of monopoles
\cite{Dirac, goddolive, WilkGold, monSU4, monSU3sinha, monSU3corri, WuYang,
EnglWind} that arise in 3 dimensions for the YMH model with group
$SU(N)$, that we will need later. The action is:
	\be \S = N\int{\rm d}^3 x\left[
	{1\over2g^2}\tr(G_{\mu\nu}G_{\mu\nu})
	+{1\over2}\tr(\D_\mu \phi\D_\mu\phi)
		+{\lambda\over N}(\tr(\phi^2)-\mu^2)^2\right] \ee.

First of all one introduces what are called {\it point monopoles}, 
singular configurations satisfying the equations of YMH plus the
requirement of minimum action:
	\be \D_\mu G_{\mu\nu} = 0 \qquad
	    \dd {\phi^a} V(\phi) =0 \qquad
	    \D_\mu \phi^a =0 \ee
The last two equations impose that one gets everywhere zero contribution to
the action from the Higgs sector, so that we are in what is called
{\it Higgs vacuum}. Some gauge fields have to be zero and the others
allow for nontrivial solutions localized near isolated points. 

\smallskip 
To see this it is useful to choose the abelian ``unitary'' gauge 
$\phi\in \h$ (with \h{} the Cartan subalgebra). 
We will write $\phi = \ph\cdot\T$, where $\T$ is the vector of $(N-1)$  
commuting generators of \h{} in some faithful representation 
and $\ph$ is a vector in $\R^{N-1}$.

In this gauge the above equations imply that the Higgs field 
is constant in all space $\phi(x)=\phi_\infty$ and that ${A}_\mu$ 
has nonzero components only in the algebra of the {\it little group} 
of $\phi_\infty$, i.e. only if $[A,\phi_\infty]=0$. 

For $\phi_\infty$ with generic eigenvalues (all different) the little
group is simply $U(1)^{N-1}$, so that only the abelian, Cartan, gauge fields
survive in the vacuum.

Here $\phi_\infty$ induces perturbatively a mass term 
for each gauge field that is charged with respect to it. 
For each root $\a$ of SU(N) the mass of the charged $A^\pm_\a$
fields is $m_{W_\al}=g|\ph_\infty\cdot\a|$. 
At the same time the Cartan gauge fields decouple completely 
from the Higgs fields.

As $V(\phi)$ is flat in all gauge directions, the $V'(\phi)=0$
constraint only fixes the modulus of the vacuum Higgs field: 
$\tr\phi^2_\infty=\mu^2$.

We will assume at this point a vacuum $\phi_\infty$ with all different
eigenvalues. This is a gauge invariant statement and we will see that
this vacuum configurations should be preferred by the system itself, in
the large $N$ limit. 
It could also be imposed by some gauge invariant external source.

{\bf \small (checked to here)}

\subsection{Point monopoles}
\smallskip 
The $G_{\mu\nu}$ field equation now allows for abelian
$U(1)^{N-1}$ solutions with a singular Dirac string \cite{Dirac}: 
        \be &&A_\mu=\T\cdot\q D_\mu  \label{Aunitary}
		\qquad\qquad D_\mu=(1-z/r)\d_\mu\tg^{-1}(y/x)\\
	    &&G_{\mu\nu} = g\T\cdot\q {\epsilon_{\mu\nu\lambda}x_\lambda
		\over |x|^3} \qquad \q \in \R^{N-1}.\0\ee
$\q$ is the nonabelian charge of the monopole, but as we are in the
unitary gauge, it belongs to $U(1)^{N-1}$: it is for now an arbitrary
vector in $\R^{N-1}$.

As $z\rightarrow-\infty$ we have the asymptotic form: 
   \be A_\mu = 2\T\cdot\q\d_\mu\Phi
\qquad(0\leq\Phi=\tg^{-1}(y/x)\leq2\pi).\ee

\smallskip
Observing the phase of a loop around the string we get a realization of 
${\pi_1}({U(1)^{N-1}})$ and obtain the admissible monopoles: 
the generalized Dirac condition
        \be \e^{4\pi i \q\cdot \T} = \id.  \label{Dirac}\ee

This condition restricts the possible charges $\q$ to belong to a {\it
lattice} in $\R^{N-1}$, in fact $\q$ has to be reciprocal to each 
weight of the representation chosen for the $\T$: for every weight $\m_i$, 
        \be \q \cdot \m_i ={n_i\over2} \qquad n \in \Z.\ee
 
The lattice of charges depends thus on the representation chosen for 
the $\T$'s, calling in the game also the global properties of the
representation of the gauge group.

It can be more or less dense depending on the modulus of the highest 
weight of the representation.
\medskip

One now introduces the (dual) {\it co-roots}, $\a^*_i = \a_i/\a_i\cdot\a_i$, 
(where $\a_i$ are the simple roots). For each weight they satisfy the relation
$\m\cdot\a^*=n/2$, so that they are reciprocal to the weight lattice.
The coroot system defines what \cite{goddolive} have called the 
{\it dual group}. 
For $SU(N)$ the dual group is isomorphic to it, denoted $SU ^*(N)$. Moreover for
roots normalized to unity the coroot lattice coincides with the 
root lattice.

An immediate consequence of the relation with the weight lattice is
that the coroots (and also their multiples) are always between 
the possible magnetic charges $\q$.
Monopoles in the adjoint representation of the dual group are thus
always present.

Usually one studies the simplest cases of the fundamental 
and adjoint representations, for the gauge field.

\smallskip
For $\T$ in the fundamental representation ($\T={1\over2}\lamb$) 
the weights are the fundamental ones and the
magnetic lattice is just the coroot lattice
        \be \q = \sum_{i=1...N-1} n_i \a_i^* \qquad n_i \in \Z\ee

The monopoles of minimum charge transform in this adjoint
representation.

The picture represent the coroot lattice for SU$ ^*$(3), its generators as
black circles. 
The small triangles represent the fundamental monopoles that
arise for adjoint gauge generators\footnote{
In fact an other case mentioned in the literature \cite{DasWadia} is
that of gauge variables with generators in the adjoint representation.
The nonzero weights are in this case the simple roots,
so that the reciprocal lattice coincides with the weight 
lattice of the dual group ($\m^*\cdot\a=n/2$): 
	\be \q = \sum_{i=1...N-1} n_i \m^*_i \qquad n_i \in\Z.\ee
One can say that the minimum charge monopoles transform now in the
fundamental representation of the dual group. They are shown in the
picture as small triangles. The monopoles that arise in this
case include also the previous adjoint charges as combinations of
minimal monopoles, although the lattice generated by them is not shown.}.
There is thus a nice duality: for gauge variables with fundamental
(adjoint) generators, the minimum monopoles transform in the adjoint
(fundamental) representation of the dual group.

\FIGURE{\epsfig{file=lattice3.eps}}

\medskip
A remark is due for the Weyl group which acts in the 
weight lattice and thus also on the magnetic charges. 
It is generated by the reflection with respect to the planes 
orthogonal to the roots, and sends every lattice that we have 
considered into itself. 
The action can be seen as a reflection also on the magnetic charges. 

Seen on the Cartan generators, it simply exchanges 
the diagonal entries (as can be easily seen in the (overcomplete) 
basis $2 (T_{ij})_{kl}=(\delta_{ik}\delta_{jl}-\delta_{il}\delta_{jk})=
\left(\hbox{\tiny\begin{tabular}{ccccc}..\\&1\\&&..\\&&&-1\\
&&&&..\end{tabular}}\right)$).

On the fields, any Weyl action is equivalent to a {\it global} gauge
rotation with respect to the generator $E_\al+E_{-\al}$, where $\al$ is
the relative root, and this shows that monopole configurations related by
Weyl symmetry are gauge equivalent. 

\medskip
This has implications on the type of monopoles: for the usual case of
gauge variables in the fundamental representation the minimum charge
monopoles are all related by Weyl transformations. 

In the adjoint representation instead monopoles are classified by the dual
fundamental weights which are not all Weyl-equivalent: they are divided in
classes according to the $N-1$ (nontrivial) elements of 
$\pi_1(SU(N)/\Z_N)$. Weyl reflections only act within these classes. 

An example for the dual $SU(3)$ is shown in the picture, the fundamental
co-weights are represented by the the small triangles, 
the right pointing giving the [3] representation, the left the [3$ ^*$].
The Weyl transformations are the reflections with respect to the 
long-dashed lines. 

One sees that the fundamental monopoles come in two triplets
invariant under Weyl, while the usual adjoint monopoles come in a 
whole sextet.

\subsection{Regular monopoles}

One notices first from (\ref{Dirac}) that if the $U^{N-1}(1)$ path 
$\e^{2\q\cdot\T\Phi}$ ($0<\Phi<2\pi$) can be continuously gauged away 
{\it in SU(N)} to the identity, then the Dirac string will decrease 
of intensity and disappear. 
The gauge transformation needed to do that is necessarily noncostant, 
so that one will end up with a noncostant Higgs field. 
If this can be done, the point monopole is the basis for a regular 
one with finite energy. 
It is clear that there is a great freedom to construct these regular
solutions of the equations of motion.

The {\it spherically symmetric} solutions have been studied in detail
in \cite{WilkGold} with a classification of all the possible
magnetic charges from which one can determine a finite action configuration.

The charges which admit spherical solutions are given by $\q = \q'-\q''$
where $\q'$ and $\q''$ are the roots of two embeddings of SU(2) in
SU(N) and $\q''$ must also be in the little group.
One SU(2) $\q'$ is needed to rotate the Higgs field to a 
radial gauge, and the other is a remaining freedom to define the
spherical gauge configuration. 

Since the factor in (\ref{Dirac}) is a loop in the chosen
representation of $SU(N)$, the process is clearly possible 
for any charge if and only if the generators are in a faithful
representation like the fundamental one\footnote{In the case for
example of the adjoint representation, there are $N$ inequivalent
loops which join the identity to the elements of the center of
$SU(N)$. So, for $N-1$ kind of point monopoles, the string is
impossible to remove and they are genuine Dirac monopoles.  An other
accident of this case is that the minimum charge monopoles are in this
case N(N-1) (the weights of all fundamental representations), while
the nontrivial elements of the center $\Z_N$ are N-1. The minimum
charges can thus be divided in (N-1) sets of N elements, and the Weyl
group acts only within each set. 

All this fundamental monopoles, associated to the nontrivial paths in 
SU(N) around the Dirac string, cannot be made regular.}.

\bigskip
The Weyl group relates monopoles of different charges by a global gauge
transformation. This is a difference with the abelian case where different
charges classify gauge inequivalent monopoles.
This peculiarity of non-abelian theories follows mainly from the
simple-connectedness of SU(N), but also from the fact that colored 
flux lines are not gauge invariant (but covariant) and can thus be 
deformed or changed of color by gauge transformation. 

\bigskip
Another important notice is that the charge quantization condition
(\ref{Dirac}) does not depend on the vacuum Higgs field $\phi_\infty$,
and this is an advantage of the unitary gauge.

Regular monopoles instead do depend on the $\phi_\infty$ boundary
condition, because they are constructed transforming
the Dirac string into a varying Higgs field.

As we are going to assume $\phi_\infty$ to have all different
eigenvalues, the charge $\q''$, belonging to the little group, can
only be zero, so that the magnetic charges of spherically
symmetric monopoles will coincide with $\q$ roots of SU(2)
embeddings\footnote{Monopoles of higher charge, empty sites in the
picture, are not realizable as single three dimensional spherical
configurations, although the topological argument 
indicates the existence of some extrema of the action. 
Some can be constructed as multimonopole-like configurations which 
possess discrete symmetry groups, with stability given by the Higgs 
attraction.  Such configurations have been found to exist with
tetrahedral, octahedral (but not icosahedral symmetry, for a late
reference, see \cite{TetraCubic} and therein).}.
The figure shows them for $SU(3)$, for $N>3$ the pattern is much more
complicated.

\FIGURE{\epsfig{file=lattice.eps}}

Only the first two cases are possible for nondegenerate $\phi_\infty$.
For this cases the solution are well-known 't Hooft-Polyakov
monopoles: 

     \be \hat A_\mu({\bf r},\q)&=& T_q D_\mu
	+ K(m_W{\bf r})\left[E_\q e^{-i\phi}(i \hat \theta + \hat \phi)_\mu
			 + E_{-\q}e^{i\phi}(-i\hat\theta + \hat\phi)_\mu\right]\0\\
	    \hat\phi({\bf r},\q) &=& \ph_\infty\cdot(\T-\q T_q)
		+H(m_H{\bf r})(\ph_\infty\cdot\q)T_q\label{monconfig}\ee

$T_q=\q\cdot\T$ with some $E_\q$ and $E_{-\q}$, give the SU(2)
 subalgebra.

Changing to a regular gauge one can ``smear out''the string in $D_\mu$ 
and leave an isolated singularity at the origin. 

\subsubsection{Masses}\label{masses}
In the last expressions, $m_W$ is the mass of the two charged gauge
bosons in the chosen sector of the monopole, function of the Higgs
vacuum: $m_W = g |\ph^\infty\cdot\q|=g|\phi^\infty_i-\phi^\infty_j|$;
$m_H$ is instead the mass of the Higgs field "in the sector of $T_q$":
$m_H^2=2{\lm\over N}(\phi^\infty_i-\phi^\infty_j)^2$.

They regulate the exponential decay of the massive field components
of the monopole as well as its total classical action.
We have in fact       
	\be \S_{cl} = N |\q| {m_W(\q)\over g^2} C({\lm\over N g^2}). \ee

The function $C$ is found \cite{BPS} to approach the value $4\pi$ when
the argument goes to zero, which for $SU(2)$ is called $BPS$ limit. 


Considering the $N(N-1)$ unit charge monopoles, we see that, fixed 
$\phi^\infty$, their pseudo-mass ranges in the interval
$0\leq \S_{cl}\leq 4\pi N \mu/g$. 

\medskip
We see that most of the properties of the various objects are ruled
by the $m_W$ of the gauge bosons in the relative $SU(2)$ sector, so that
all depends on the set of Higgs vacuum eigenvalues $\phi^\infty_i$.


\medskip
The constraint $|\phi^\infty|=\mu\simeq O(N^{1/2})$ has important 
consequences on the set of masses that are present in the model in the
large $N$ limit, because, as we will see, necessarily there are masses that
become small at least as $O(1/N)$. We will also find masses of order
$O(1/N^2)$.


\bigskip
All the monopoles of charge greater than $1$, have higher mass 
proportional to their charge and they are expected to
dissociate into smaller constituents. Hence their contribution to the
infrared region is negligible and one can safely discard them.

\bigskip
We will instead use approximate solutions built by superimposing an
arbitrary number of minimal monopoles at large distances.

They are constructed easily in the unitary gauge and are then 
regularized by means of a procedure similar to the one for the
single monopole. We just need the proof of existence of the gauge
transformation needed to do that \cite{JaffeTau}, because we will work in
the unitary gauge.

	\be \ba{l}A_\mu^{(n)}(x)\equiv
		\sum_{a=1}^n \hat A_\mu(x-x_a,\q_a)\\
	     \phi^{(n)}(x)\equiv
		\sum_{a=1}^n (\hat\phi(x-x_a,\q_a)-\phi^\infty)+\phi^\infty
		\ea\label{multimon}\ee

They depend only on the parameters of the $n$ single monopoles
$\{x_a,\q_a\}$ and are good solution to the equations of motion
for distances much bigger than the monopole sizes. This
approximation is called {\it dilute gas}.

\medskip
The interaction of monopoles simplifies drastically
for large distances compared to the monopole size and remains
function of the relative distances only.

In fact the action for such dilute configurations is found to be 
approximated by the self-action of each monopole plus a
Coulomb-like monopole-monopole interaction term \cite{Poly75}.


\section{Quantum fluctuations}

Coming to the much more ambitious program of quantizing the theory, 
the only feasible way is for now the semiclassical expansion, leaving 
apart the S-duality which solves the supersymmetric theories, 
or the first order (BF) formalism, which also falls back to
semiclassical quantization. 

If semiclassical it is, it should show the correct result. 

However even this treatment at one loop order is a nontrivial task,
because involves the calculation of functional determinants in 
classical backgrounds.

There are some simplifications in the BPS limit \cite{BPS,JaffeTau} of
supersymmetric theories, because the three-dimensional configurations
represent four dimensional selfdual backgrounds and there are useful
relations between fermion and boson determinants in this case,
but we remind that the physics in this limit is drastically different.
The main reason for this is that the Higgs field is massless and thus
gives a further long range interaction. We thus want to consider $\lm>0$.

\medskip
A major progress was made by Polyakov. In analogy with other simpler
models he applied \cite{Poly77} the semiclassical quantization to
the $SU(2)$ monopole solutions and carried the program to evaluate 
the Wilson loop resumming the semiclassical expansion.

His treatment, for the compact QED, shows that the area law for the
Wilson loop emerges from this semiclassical treatment precisely because
of the condensation of monopoles.

Later in \cite{DasWadia} Das and Wadia have reached the same conclusion
for the problem with $SU(3)$ gauge group.
Recently also for the case of {\it pure} $YM_3$ they have argued that
confinement arises from the gas of monopoles, using nonperturbative
results from the theory of the three dimensional Coulomb gas
\cite{DasWadia94}.

\medskip
So we have, at our disposal, the minima of the classical action that are
strongly supposed to give the main contribution to the functional
integral, and taking into account the gaussian fluctuations 
around them, one can introduce a semiclassical sum.

We will see that important nonperturbative features of the model
are reproduced by this approach.

\subsection{Semiclassical program}

In \cite{Poly77, DasWadia} the semiclassical quantization of a system with
monopoles is approached through a grand canonical ensemble of magnetic
particles. The sum on all configurations gives the partition function
in a nonperturbative way, and after a generalized Poisson transform, 
a saddle point technique can be applied. 

The gauge fields should be integrated perturbatively at one loop, taking
into account the regular monopole backgrounds as nontrivial minima of
the action.

\medskip
In this approach many approximations are to be taken carefullly:

First, the gas of monopoles is assumed dilute, thus considering
just the coulomb part of the monopole-monopole interaction and
simplifying the classical and one-loop effective action.

After the generalized Poisson\& Sine-Gordon transform has been used, 
usually one is limited to the minimum charge monopoles, assuming the 
higher ones dissociate rapidly\footnote{%
One could try to consider also the higher charges. From the Wilkinson
Goldhaber analysis, that spherical monopoles have limited charge 
(for limited N, $|\q|\leq N-1$ ). Moreover their mass grows linearly
with the charge while their size gets linearly shrinked.}.

\subsection{Grand canonical ensemble} \label{gas}

The partition function of {\it pure} YM in 2+1 dimensions is 
transformed into the sum on all configurations made of any number of
monopoles of charges $\{\q_a\}$ and locations $\{x_a\}$. For $SU(2)$
it is:
	\be Z = \sum_{n=0}^\infty {\xi^n\over n!} Q_n \0\ee
	\be Q_n = \sum_{\{\q_1\cdots\q_n\}}\int\prod_a {\rm d}^3x_a\,
	 \e^{{-2\pi\over g^2}\sum_{a\neq b}{q_aq_b\over|x_a-x_b|}}
		\qquad (q_a=\pm1,\pm2,\ldots)\0\ee 
where $\xi$ is the classical and one loop contribution to 
the action of the one monopole configuration \cite{Poly77}.

For $SU(N)$ this expression is no longer valid because the weight of each
monopole depends on the charge, $\xi=\xi(\q)$. Hence one has:
	\be Z = \sum_{n=0}^\infty {1\over n!} Q_n \label{SC}\ee
	\be Q_n = \sum_{\{\q_1\cdots\q_n\}}\int\prod_{a=1}^n {\rm d}^3x_a\,\xi(\q_a)
	\cdot \e^{{-2\pi\over g^2}\sum_{a\neq b}{\q_a\cdot\q_b\over|x_a-x_b|}}
		\qquad (\q_a\ in\ the\ root\ lattice)\ee 	
\bigskip

The partition function $Z$ can be nicely reexpressed as a functional integral
using the Sine-Gordon transform \cite{PolyBook}: 
	\be Z = \int \D\ch \e^{-{g^2\over32\pi^2}\int{\rm d}^3x\,
		\left[(\d\ch)^2-\sum_iM^2(\q_i)e^{i\q_i\cdot\ch}\right]}. \ee
The mass $M^2$ comes from the weight $\xi$: $M^2 = 32\pi^2\xi/g^2$.
$\ch$ is a $N-1$ components scalar field whose propagator is just 
the coulomb potential, and the sum on $i$ is the sum on all
the possible magnetic charges of one monopole (the magnetic lattice).
The last term in the integral is the functional generator of the 
multi-charge configurations. 
For the symmetry of the minimum charges, the roots $\q_i=-\q_{-i}$,
the last term becomes a cosine.

\medskip
This representation of the coulomb gas has to be understood in a
perturbative sense, because it is just the perturbative expansion
which reproduces, diagram by diagram, the dilute gas. In this spirit
the $\ch$ field configurations have to vanish at infinity, although
there is formally an infinite 'zero-point' energy of the vacuum
$\ch=0$, which disappears with the normalization.

\subsection{Wilson loop}

One can then succeed to evaluate gauge invariant operators like the 
Wilson loop $W_C$:
      \be W_C &=& {1\over N} \< \Tr\,\e^{i\int_CA\cdot{\rm d}{\bf x}}\>\ee
for any contour C in three dimensional space-time.

If we take into account the form of $A_\mu$ in the unitary gauge
(\ref{Aunitary}) for each monopole of charge $\q_a$ and location $x_a$, 
(and after using the Stokes theorem) we can rewrite the Wilson loop as an
external source for the configuration of charges $\{\q_a\}$:
	\be W_C &=& {1\over N} \< \Tr\,\e^{i \T\cdot\sum_a \q_a\int_S
                  {\rm d}^2\sigma_\mu(y){(x_a-y)_\mu\over|x_a-y|^3}}\>
	\qquad\eta(x)=\int{\rm d}^2\sigma_\mu(y){(x-y)_\mu\over|x-y|^3}\0\\
	&=& {1\over N} \< \Tr\,\e^{i\T\cdot\sum_a\q_a\eta(x_a)}\>\ee 

\FIGURE{\epsfig{file=eta.eps}}
Instead of the flux through the loop of the magnetic field produced 
by the $\q_a$ charges, one thinks to a potential ($\eta(x)$, in the picture) 
which acts on the charges $\q_a$ at points $x_a$, produced by a dipole 
layer on the surface S spanning the loop\footnote{The dipole density
is unitary and in direction of $\di^2\sigma_\mu$ orthogonal to the
surface. The potential is in practice the solid angle of the loop seen
by the charge at $x_a$.}. The problem of evaluating the Wilson
loop in the functional integral is reduced to the average of the
canonical ensemble under the action of this external potential.

In the limit of very large loop this potential is constant on the two sides 
leaving just the discontinuity, which is of crucial importance. 

From the color trace the potential takes each direction along the fundamental weights $\m_\al$:
     \be W_C &=& {1\over N} \< \Tr\,\e^{i\T\cdot\sum_a\q_a\eta(x_a)}\>
	= {1\over N} \sum_\al\<\e^{i\sum_a\m_\al\cdot\q_a \eta(x_a)}\>\ee

\medskip
It is straightforward to evaluate this operator in the coulomb gas ensemble
(\ref{SC}), and the result is, after a shift in the $\ch$ field: 
    \be W_C = {1\over ZN}\sum_\al \int\D\ch\e^{-{g^2\over32\pi^2}
	\int{\rm d}^3x \,\left[
	(\d\ch-\m_\al\d\eta)^2-\sum_iM^2(\q_i)\e^{i\q_i\cdot\ch}\right]}
				\label{Wchieta}\ee

\subsection{Saddle point solution}

The nongaussian integral above is solvable by saddle point,
remembering that $M^2$ is asymptotically small.

The saddle point is given by the equations, one for each $\m_\al$: 
	\be -\d^2\ch_\al + \m_\al\d^2\eta = 
		- \sum_{half\ i}M^2(\q_i) \q_i\sin(\q_i\cdot\ch_\al)\ee

(The sum on $i$ runs now only on one half of the magnetic lattice 
because the negative symmetric part is included in the sine).

The term $\m_\al\d^2\eta$ imposes a discontinuity of the solution
in 'internal' direction $\m_\al$ and together with the constraint
of perturbativeness $\ch(\pm\infty)=0$, this leaves only the solutions
of constant direction $\ch(x) = \m_\al \chi_\al(x)$ (no sum). 

Taking the product with $\m_\al$, one gets the scalar equations 
	\be (\d^2\chi_\al-\d^2\eta){N-1\over 2N}= \sum_{half\ i}
	       M^2(\q_i)\m_\al\cdot\q_i\sin(\q_i\cdot\m_\al\chi_\al)\ee
(where we used the fact that for the weights of the fundamental repr. 
$\m_\al^2={N-1\over2N}$). 

\medskip
At this point one has to specify which magnetic charges $\q_i$ to use, 
and we see from the Sine-Gordon representation that higher charge
monopoles give rise to a generalized Sine-Gordon potential which
is of lower magnitude, even if it has shorter periods.
This means that higher charges only perturbate the potential given by
the minimal charges. 

\medskip
We limit then the sum over $i$ to the minimum 
magnetic charges, i.e. the co-roots of the first picture.

We have to evaluate the scalar product $\m_\al\cdot\q_i$ of a
fundamental weight with the adjoint weights.

To this aim we remember that the roots $\q_i$ are $N(N-1)$, and that  
given a fundamental weight $\m_\al$, $(N-1)(N-2)$ of them are orthogonal
to it, while $N-1$ have scalar product $1/2$ and the others $(N-1)$ 
(negative symmetric) have scalar product $-1/2$. 

It is then sufficient to limit the sum to the $(N-1)$ cases all giving 
result $1/2$ for the scalar product: 
     \be \d^2\chi_\al=\d^2\eta +{2N\over N-1}{1\over2}\sin(\chi_\al/2)
			(N-1)\bar M^2_\al\label{scaleq}\ee
where we have introduced the {\it averaged} $\bar M^2$ given by
	\be \bar M^2_\al={1\over (N-1)}
		\sum_{\q_i:\q_i\cdot\m_\al=1/2} M^2(\q_i).\ee

It is this quantity which carries information on the physics that we
obtain in the large $N$ limit. Each $M^2(\q)$ is, in Coulomb gas
language, the {\it fugacity} of the monopole species $\q$. $\bar M^2_\al$
is then directly related to the average density of monopoles, which is
strongly believed to be the order parameter for confinement.

We note also that many monopoles are mutually neutral and that $\bar M^2_\al$
is the average in the $(N-1)$ sectors that are not orthogonal to $\m_\al$.
To evaluate the Wilson loop we should also average on $\m_\al$, finally.
		
\medskip
The solution of (\ref{scaleq}) is known explicitly:
\be
    \chi_\al(x)=
  \left\{   \ba{ll}
        4\tg^{-1}\e^{-\bar M\sqrt{N\over2}x}~~~~~~~~~~& x>0\\
      - 4\tg^{-1}\e^{ \bar M\sqrt{N\over2}x}          & x<0\ea  \right.
\ee
which consists of two parts of a Sine-Gordon soliton. We remark that
this soliton is permitted only because the discontinuity 
allows the field to vanish at infinity. Otherwise there could be an
infinity of other classical solutions.

Inserting this into eq. (\ref{Wchieta}), it gives the estimate for the
Wilson loop:
	\be W_C &=&{1\over N}\sum_\al\exp\left\{-{g^2\over32\pi^2}
		\int{\rm d}^3x\left[{N-1\over2N}(\chi_\al'-\eta')^2
			-\bar M^2(N-1)(\cos(\chi_\al/2)-1)\right]\right\}\0\\
		&\simeq& {1\over N}\sum_\al\e^{-\sigma_\al A}\ee
with string tensions $\sigma_\al$ of
  \be\sigma_\al={g^2 \bar M\over32N\pi^2} {N-1\over2}\sqrt{{2\over N}}
	   = {g\over 8\pi} {N-1\over N}\sqrt{\bar\xi\over N}\ee

This shows that confinement of quarks exists in this theory for
generic values of the coupling constants and for finite
$N$, while to extract the behavior with large $N$ it is necessary to perform
the average for $\bar M$ and then for the Wilson loop
${1\over N-1}\sum_\al\e^{-\sigma_\al A}$.

\bigskip
The $N$ dependence of $M$, that is of $\xi$, is not known explicitly.
As in \cite{Poly77}, $\xi$ is the one loop partition function in a
single monopole background, before the integration of the zero mode
translation coordinate:
	\be \xi(\q) &=& N^{9/2}m_W^3\left({m_W^{3/2}(\q)\over g^3}\right)
			A(\lm/ g^2) \e^{-N{m_W(\q)\over g^2}C(\lm/Ng^2)}\ee
The condition of validity of the saddle point approximation, which
represents the low Debye density, is $\bar \xi<<1$. It can be seen to
hold for finite $N$ from the above expression, where the exponential
vanishes asymptotically.
In the limit of $N$ large one needs a more precise estimate 
on the whole average $\bar\xi$. This we will attempt in the next sections.

%
%

\setcounter{footnote}{0}

\subsection{Determinant}

In the previous sections, $\xi(\q)$ plays the part of 
the statistical-quantum weight of a particular background 
configuration, so that the higher it is, the higher is the 
importance of that particular configuration, 
although it may have large action. 

\smallskip
Up to now we assumed $\xi$ to be some fixed quantity. Now, in order to draw
some conclusion about the string tension $\sigma$, we need to find something
more precise on it.

The evaluation of $\xi$ is the problem of calculating the functional
determinant of the fluctuating fields around the one monopole solution.
It would be a hopeless problem to calculate it exactly in an arbitrary  
external field, as it is equivalent to the solution of a Schroedinger or
Dirac equation in an external potential\footnote{
In \cite{Oleszcz} the expression has calculated numerically with the
heat kernel method for $SU(2)$, but the estimate of the behavior with $N$
has yet to be done.}.

We will try to extract some information from the high $N$ analysis
of the problem.

The idea comes from the fact that after gauge fixing, and better in
the unitary gauge, the Higgs field has only $N$ components, while its
effective action, upon integration of the gauge sector, is of order $N^2$.
Hence the saddle point should be applicable, and the Higgs field is a
semiclassical quantity with respect to $1/N$, which acts like $\hbar\to0$
to suppress its fluctuations. 

\smallskip
We will treat the fields in one loop approximation around the one monopole
configuration $\hat\Amu, \hat\phi$ adding the fluctuating fields
$\amu,\varphi$, so that:  $\Amu = \hat\Amu+\amu$, $\phi = \hat\phi+\varphi$. 

\medskip
In doing so, we are faced with the problem of gauge fixing, because
there are zero modes of the action. The gauge invariance involves
the total field $(\hat\Amu+\amu,\hat\phi+\varphi)$, and one can split
the gauge variation between the fluctuating and the background fields in
an arbitrary manner.

Among the possible (infinite) choices, one can assign the whole field
variation either to the fluctuations or to the background. The latter
choice is of little or no utility, the first, instead, is quite convenient
in that it keeps away the gauge invariance problem from the background fields.

So we will keep the background fields in some fixed gauge, and consider
the gauge group as acting on the sole fluctuations:
\be	&&\delta_{gauge} (\hat\Amu,\hat\phi)=0 \0\\
	&&\delta_{gauge} (\amu,\varphi)=\delta_{(\Amu,\phi)}(\amu,\varphi)=
		(\D_\mu\al,-ig[\phi,\al]).\ee

The gauge for the classical fields is left for now unspecified, even
the radial gauge satisfy automatically the background gauge. In the next
section instead we will choose for all fields the unitary gauge.

\bigskip
The partition function in a one-monopole background, $\xi$, is a gauge
invariant object but not gauge independent, at least in principle.
Every gauge fixing could provide different physical insight, as happens
already with spontaneous symmetry breaking.

According to 't Hooft and Polyakov $\xi$ it is better calculated in the
so called {\it background, "natural", gauge}, for two reasons: one is
the presence of zero modes, to appear in a moment, for the action of
fluctuations, and they are best treated in the background gauge;
the second is that candidates with opposite features, the unitary gauges,
usually addressed to be non-renormalizable, can be cured 
only introducing a nonpolynomial term in the action.

This can be seen from the Vandermonde determinant 
appearing in the measure of $\vph$ after elimination of the
($SU(N)$) 'angular' part of $\varphi$: 
$\D\varphi\rightarrow\D\vph\prod_x\Delta^2(\vph(x))$.

There are two ways to deal with this $\Delta^2$; it can be reabsorbed
in the measure by a (nonlinear but nonsingular) change of $\varphi_i$
variables, but it yields a nonlinear model on a singular curved target
space of the kind (for $SU(2)$ for example):
	\be \L= \varphi^{-{4\over3}}(\d_\mu\varphi)^2\ee

Alternatively, thinking to a lattice, one can exponentiate the
determinant and convert it in the divergent logarithmic potential:
$\delta(0) \sum_{i\neq j}\int\log|\phi_i-\phi_j|\dtre x$, but the
continuum limit seems problematic due to the ultralocal nature
of $\Delta^2$. This procedure can be justified with care starting from
the $R_\xi$ kind of gauges ['tH,V] \cite{Sher} or with other
kind of regular gauge fixing.   

\medskip
I can note however, that this divergent extra potential,
being exponentiated without the help of ghost fields, carries an $\hbar$
factor, and thus already is a {\it quantum} correction to the action.

As the theory is renormalizable for every gauge $R_\xi$, in the $\phi$
effective action there should be, then, a divergent term from the massive
gauge fields which cancels it.
In effect it is what we find after the one loop analysis. 

{\it Hence for the lare $N$ saddle point analysis on the effective
potential, there should be no serious problem from this term. }

We will adopt in the following paragraph the background gauge and in
the next the unitary one.

\subsubsection{Background Gauge}

Here we take, as gauge fixing: 
	\be F_\kappa = \hat\Dmu\amu -i\kappa g[\hat\phi,\varphi]=B(x)
			\label{gf}\ee
In the limit $\kappa\rightarrow\infty$ the second term reproduces the
unitary gauge. This is a variant of the pure background gauge (that has
$B=0$, $\kappa=1$), already considered by \cite{tHooft} or of the
"natural gauge" that appears in \cite{Poly77}. It has a long
history in the literature, due to its double features of
renormalizability together with massive ghost fields. 

\smallskip
The usual gaussian averaging of the gauge fixing 
$\delta(F_k-B)$, to get a quadratic term in the action, is in some
contrast with the high $N$ analysis because removes the gauge fixing 
and leaves $N^2$ degrees of freedom. On the other hand 
fixing strictly the gauge in the case $k\neq1$ presents some subtleties. 

\medskip
We will denote together the fluctuating fields $\amu(x)$ and $\varphi(x)$
with $\Phi = \bv\amu\\\varphi\ev$ and the full quadratic action will be, 
in matrix form, $\S_{quadr}=\null^t\Phi\cM\Phi$. Scalar product
$\Phi\cdot\Phi$ will imply integration on space-time.

About the action we just need to say, for now\footnote{
	Explicitly it is:
	\be\S_{quadr} = N\tr\int \left[([\hat\Dmu,\anu]-[\hat\Dnu,\amu])^2
			+{g\over2}\hat G_{\mu\nu} [\amu,\anu] 
			+( g [\amu,\hat\phi+\varphi]+[\hat\Dmu,\varphi])^2
			+ \null^t\varphi V''(\varphi) \varphi \right]
				\0\ee
	Together with the proper mass term for gauge fields,
	${1\over2}Ng^2[\hat\phi,\amu]^2$, there is also the bilinear mixing 
	$N([\amu,\varphi][\hat\Dmu,\hat\phi]+[\amu,\hat\phi][\hat\Dmu,\varphi])$.
}, that it is annihilated by the following zero modes: 
       \be gauge &&\Phi_0(\al)= \bv\amu(\al)\\\varphi(\al)\ev =
			\bv\hat\Dmu\al\\-ig[\hat\phi,\al]\ev\label{gzm}\\
     translation &&~~\Phi_0^{(i)}~=~\bv\bar\amu^{(i)}\\\bar\varphi^{(i)}\ev
	~=~ \bv{1\over g}\hat G_{\mu i}\\\hat D_i\hat\phi\ev
		\label{tzm}\ee

The complete field $\Phi$ can thus be decomposed in zero modes plus
nonzero ones, $\Phi_n$, eigenfunctions of the action:
$$\Phi = R_i\Phi_0^{(i)} + \Phi_0(\al) + \xi_n\Phi_n $$

The translational modes (\ref{tzm}) are written in a gauge so that they
satisfy the gauge-fixing (\ref{gf}) with $\kappa=1$, $B=0$. 
This is useful because they are orthogonal to the gauge modes.
They are normalized to

\be |\Phi_0^{(i)}|^2=\cN_i=N\tr\int{1\over g^2}\hat G_{i\mu}^2
			+ (\hat D_i\hat\phi)^2.\label{normtzm}\ee
				
All these zero modes are treated with the standard Faddeev-Popov method
to extract the integration on collective coordinates $R_i$, $\al(x)$:
\be 1 = \Det_{ij}\left[\dd{R_j}\Phi\cdot\Phi_0^{(i)} \right]
		\Det\left[{\delta\over\delta\al(x)}F_k(\Phi(y))\right]
				\cdot
		\int \dtre R\D\al\,\delta(F_k-B)\prod_{(i)} 
			\delta\left(\Phi\cdot\Phi_0^{(i)}\right).\0\ee
Insertion of this unity in the functional integral replaces 
the flat fluctuations with the right variables $R_i$, $\al(x)$ 
and gives the two determinants. The first $\Det$ gives
$\prod_i\cN_i^{1/2}$, the second is the Faddeev-Popov determinant

    \be \Det\left[{\delta\over\delta\al(x)}F_k(\Phi(y))\right]=
	\Det[\cM_{FP}]=
	\Det\left[\hat\Dmu\Dmu -\kappa g^2[\hat\phi,[\phi,\cdot]]\right]\ee
After elimination of the first delta, the remaining fluctuations
represent the functional integral restricted the non translational modes:
\be Z = \e^{-\S_{cl}}\int\dtre R \prod_i\cN_i^{1/2} \int\tilde\D\Phi 
		\delta\left(F_k(\Phi)-B\right)\Det[\cM_{FP}]
			\e^{-\int\null^t\Phi\cM\Phi}\ee
The Faddeev-Popov determinant can be evaluated, in one loop approximation,
in the sole classical background fields. 

Then, as seen from (\ref{normtzm}), each factor in $\prod_i\cN_i^{1/2}$ 
is the action in a space-time direction without the potential.
Because we consider spherical monopoles only, all $\cN_i$ are equal and
	\be \cN_i={N\over3}\tr\int {1\over g^2}\hat G^2+(\hat D\hat\phi)^2\ee
They coincide with the action in the BPS limit, hence, after the
discussion of section {\bf \ref{masses}}, in the large $N$
limit we also take $\cN=N {4\pi\over3}{m_W\over g^2}$.

\bigskip
One can perform a functional integration over $B$ 
(with $\e^{-{\lm\over2}\int B^2}$) to remove the $\delta(F^k-B)$:

	\be Z &=& \int\dtre R\e^{-\S_{cl}} \cN^{3/2}
    \Det\left[\hat\Dmu\hat\Dmu-\kappa g^2[\hat\phi,[\hat\phi,\cdot]]\right]
	\int\tilde\D\Phi
		 \e^{-\int\null^t\Phi(\cM+{\lm\over2}\cM_{gf})\Phi}\0\\
	&=& \int\dtre R\e^{-\S_{cl}} \cN^{3/2}\Det
	\left[\hat\Dmu\hat\Dmu-\kappa g^2[\hat\phi,[\hat\phi,\cdot]]\right]
	\widetilde\Det^{-1/2}\left[\cM+{\lm\over2}\cM_{gf}\right]
			\label{pol}\ee
	
Here with $\null^t\Phi\cM_{gf}\Phi=F_k^2(\Phi)$ we denote the standard
gauge fixing term arising from $F_\kappa$, while with
$\tilde\D\Phi=\tilde\D a\tilde\D\varphi$, as with the determinant
$\widetilde\Det$, we integrate on translation-fixed fluctuations.

This is analogous to the formula of Polyakov \cite{Poly77}, but does not
show the explicit dependence on $\kappa$; one has to go through the
loop expansion in gauge bosons and ghosts in the external field,
and proceed to resum all contributions of generic order in $g$ if one
wants to control the limit $\kappa\to\infty$. Actually in this limit the
ghost masses become large so that they decouple, but at the same time
the coupling with gauge and Higgs also diverges, so that the resumming
is nontrivial. We will see the first order in an other way in the next
section.

\medskip
Now we proceed in a different direction, we eliminate the
$\delta(F_\kappa-B)$ by direct integration of the gauge modes.

This point of view has the advantage of showing the physical degrees
of freedom $\Phi_n$, together with the explicit dependence on $\kappa$
that we want to compare with the unitary gauge. 

Briefly, instead of $\widetilde\Det\left[\cM+{\lm\over2}\cM_{gf}\right]$ we 
get the decoupled product of two determinants
$\widetilde\Det_{gf}\left[\cM\right] 
\Det^2\left[\sqrt{\lm\over2}\cM_{FP}\right]$ plus a measure
jacobian dependent on $\kappa$.
The second determinant cancels formally the FP determinant, and $\lm/2$
disappears in the normalization. $\kappa$ of course remains in
the measure, but the limit is nonsingular

\bigskip
Let's expand as before the fields as
$\Phi = R_i\Phi_0^{(i)} + \Phi_0(\al) + \xi_n\Phi_n$. We can eliminate
the zero modes, taking into account the jacobian from $\Phi$ to $(\al,R_i)$.

We get (still ignoring contributions at more than one loop):
	\be Z = \int\dtre R\,\e^{-\S_{cl}} \prod_i\cN_i^{1/2}
	\sqrt{\Det\left[\hat\Dmu\hat\Dmu-g^2[\hat\phi,[\hat\phi,\cdot]]\right]}
	\int\D (\xi_n\Phi_n) \e^{-\xi_n^2\int\Phi_n{\cM}\Phi_n}
	\ee
The dependence on $\kappa$ and $B$ seems disappeared, but 
the eigenfunctions $\Phi_n$ have to satisfy the gauge fixing
so that they are sensitive to $\kappa$ and $B$.

Remembering that $\Phi_n$ was satisfying the natural gauge ( (\ref{gf})
$B=1$, $\kappa=0$), to pass to a choice with $B\neq0$, $\kappa\neq1$
we have to perform a gauge transformation of the $\Phi_n$,
so that they satisfy the new gauge fixing:
	\be \Phi_n \rightarrow \Phi_n + \Phi_0(\al_n).\ee

This gauge transformation is:
			\be\al_n(B,\kappa-1)=
	\left[\hat\Dmu\hat\Dmu-g^2[\hat\phi,[\hat\phi,\cdot]]\right]^{-1}
		\left(ig(\kappa-1)[\hat\phi,\varphi_n]+B\right)
				\label{gaugechange}.\ee
Notice that $\al_n(B,\kappa-1)\to\infty$ when $\kappa\to\infty$.

Because we are just mixing with components along the gauge zero modes,
the quadratic form in the action is unaffected.

The new basis $\Phi_n$ is still orthogonal to the translational modes.
For this reason we can retain the same measure of translations $\cN^{3/2}$.

However we have now eigenfunctions which are non normalized and not even
mutually orthogonal, and we have to re-normalize the measure. 

  \be \tilde\D\amu\tilde\D\varphi=\D[\xi_n\Phi_n]=\prod_n\di\xi_n
		&\rightarrow&
		\D[\xi_n\left(\Phi_n+\Phi_0(\al_n) \right)]
		= \prod_n\di\xi_n \cdot J \0\ee
	\be J= \sqrt{{\det}_{mn}\left[{\bf 1}_{mn} +\phi_0(\al_m)\phi_0(\al_n)
				\right]}\ee
	\be\e^{-\int\null^t\Phi\cM\Phi}=\e^{-\cM_{nn}\xi_n^2} &\rightarrow&
	 	\e^{-\int(\null^t\Phi+\Phi_0(\al))\cM(\Phi+\Phi_0(\al))}
			=\e^{-\cM_{nn}\xi_n^2}\0\ee

The new jacobian $J$ carries the information about the gauge dependence
on $\kappa$ of the functional integral. $J$, although a formal expression,
is explicitly function of the gauge parameter. The partition function is:
	\be \xi = \int\dtre R\e^{-\S_{cl}} \cN^{3/2} 
	\sqrt{\Det\left[\hat\Dmu\hat\Dmu-g^2[\hat\phi,[\hat\phi,\cdot]]
	\right]}\cdot J\cdot\prod_n{\rm d}\xi_n\e^{-\cM_{nn}\xi_n^2}
				\ee

For example in the case $\kappa=1$, $B\neq0$ we have
$\al_n(B,\kappa-1)=\al(B)$ and
$J= \sqrt{{\det}_{mn}\left[{\bf 1}_{mn} +\left|\phi_0(\al(B))\right|\right]}$
It does not depend explicitly on eigenfunctions, but when
$\kappa\neq1$ we have:
	\be &J=\sqrt{\det\left[
	{\bf1}_{mn}+\Phi_0\left(\al_m\right)\cdot
		    \Phi_0\left(\al_n\right)
			\right] }\\
	&\al_n(B,\kappa-1)=
	\left[\hat\Dmu\hat\Dmu-g^2[\hat\phi,[\hat\phi,\cdot]]\right]^{-1}
		\left(ig(\kappa-1)[\hat\phi,\varphi_n]+B\right).\ee

We can rewrite $J$ explicitly as a function of $\kappa$ as:

	\be J=\sqrt{\det\left[\id_{mn} + (\kappa-1)^2 
	\hat D_\mu a_\mu^{(m)}{1\over\hat\Dnu\hat\Dnu-g^2[\hat\phi,[\hat\phi,\cdot]]}
	\hat D_\mu a_\mu^{(n)}\right]}\ee
and after rescaling $(\kappa-1)^2$, we can write:
	\be &J=\sqrt{
	\det\left[\hat D_\mu a_\mu^{(m)}{1\over\hat\Dnu\hat\Dnu
-g^2[\hat\phi,[\hat\phi,\cdot]]}\hat D_\mu a_\mu^{(n)}\right]}\cdot\0\\
	&\cdot\sqrt{\det\left[\id_{mn} + {1\over(\kappa-1)^2} 
\left(\hat D_\mu a_\mu^{(m)}{1\over\hat\Dnu\hat\Dnu-g^2[\hat\phi,[\hat\phi,\cdot]]}
	\hat D_\mu a_\mu^{(n)}\right)^{-1}\right]}\ee

The conclusion that we would like to draw from the present calculation 
is that in the limit $\kappa\to\infty$ the singular behavior of the 
FP determinant in (\ref{pol}) has been canceled by the gauge bosons.

We pass directly to the unitary gauge, then, and we extract some information
about the large $N$ limit.

\subsubsection{Unitary gauge} \label{qu}

In this section we fix the gauge by requiring the Higgs field to
be diagonal.

Choosing the unitary gauge eliminates all the local gauge
invariance apart from the little group and the aforementioned discrete
Weyl group.

In the case that we consider, we assume the background Higgs field
to have all different eigenvalues and the fluctuation to be small with
respect to it, so this is precisely the case. We will verify a
posteriori whether this vacuum configuration is preferred.

\smallskip
The residual $U^{N-1}(1)$ abelian gauge invariance has to be cured in a
second moment by means of a further gauge fixing.

\medskip
In this unitary gauge the Goldstone bosons are explicitly "eaten up"
with the standard mechanism by the relative gauge fields which acquire
one polarization more together with the mass.

At the same time two things happen: first the gauge fixing requires,
through proper handling of the integration measure, the introduction of
a Faddeev-Popov determinant; second the massive gauge fields have a
Proca propagator, which carries bad behavior at large momentum.

For this last peculiarity the unitary gauge is usually addressed as
non renormalizable, because the gauge fields produce, even at one loop,
a new set of counterterms not present in the original lagrangian.

There is a lot of literature on the self-canceling of
some of these non-renormalizable divergences, starting from
\cite{LeeYan}, the remaining divergences are found (see
\cite{unitgaugeQED} and therein) to vanish on the equations of
motions, so that on-shell amplitudes do not suffer of this problem. 


We will see that the Faddeev-Popov determinant participates exactly to
render the theory manifestly renormalizable.

\medskip
This can already be inferred from the $\xi\to\infty$ limit of the $R_\xi$
gauge (for the charged gauge fields): for any value of $\xi$,
the theory is renormalizable \cite{HooVel, Hoo1, Hoo2, Lee, LeeZin}, 
so that the only counterterms needed are of the
same form of the lagrangian. The limit $\xi\to\infty$ is well defined for
the massive gauge propagator, so that assuming some suitable
regularization (Dimensional regularization is not suitable for this
scope because the divergences we have are of the kind $\delta(0)$ which
vanish identically: $\int\di^dp=0$) the cubic divergences cancel order
by order between ghosts and massive gauge fields (or, alternatively,
between the Faddeev-Popov determinant and the gauge fields). 
While the charged ghosts acquire an infinite mass, thus decoupling, 
also their coupling to the Higgs becomes large, leaving a correct
counterterm to the gauge divergences.
It follows that in the careful limit we do not expect any new
effective interactions.

A cancelation of this kind has been proven to happen in an abelian
gauge model by Appelquist and Quinn \cite{AppQui}.

\medskip
Explicitly we impose the unitary gauge by the constraint\footnote{
It is the same thing to consider the $\xi\to\infty$ limit of the
$R_\xi$ background gauges of the last section, namely $[\phi^0,\phi]=0$
or $[\hat\phi,\varphi]=0$.
}:
	\be F(\varphi) \equiv \varphi^{charged} =0\ee
It has no derivatives and acts on a field which transforms locally under
the gauge group, hence it requires the pointwise Faddeev-Popov jacobian:
	\be \left.
	    {\delta\over\delta\al(y)}F(\delta_\al\varphi(x))\right|_{F=0}=
	    [\phi,\cdot]\delta(x-y)=[\hat\phi+\varphi,\cdot]\delta(x-y)\ee
that gives the following functional determinant:
	\be \Det[\cM_{FP}]=\prod_x\prod_{i<j}
		(\hat\phi_i+\varphi_i-\hat\phi_j-\varphi_j)^2=
		\prod_x\Delta^2(\phi(x))\ee
where $\Delta(\phi)=\prod_{i<j}(\phi_i-\phi_j)$. Here a regularization
has to be implicit to make sense of the infinite product.

\medskip
However $\Det[\cM_{FP}]$ can be exponentiated without the help of ghost fields
thanks to the relation $\Det A=\e^{\Tr\log A}$ to yield an effective
potential for the Higgs field: 
	\be \e^{\sum_x \sum_{i<j}\log(\phi_i-\phi_j)}=
		\e^{\delta(0)\int\di^3x\log\Delta^2(\phi(x))}.\ee
Then, because the action is multiplied by a $-1/\hbar$ factor, we
need to multiply by a $-\hbar$ factor, so that it ends up describing
a {\it one loop} correction to the bare action (in the form of a 
repulsion of the $\phi$ eigenvalues, as in matrix models).

In three dimensions this correction has a cubic divergent coupling 
constant and a non polynomial structure.

\medskip
Together with this one loop correction, we have then to consider the
other contributions from the propagating fluctuations, namely the one loop
diagrams of gauge fields $\amu$ and diagonal Higgs field $\varphi$, to
calculate, at one loop:
  \be \xi=\cN^{3/2}\int\di^3R\int\D\varphi\int\D\amu\e^{-{1\over\hbar}\left[
     S_{cl}+S_{quadr}-\hbar\delta(0)\int\di^3x\log\Delta^2(\phi(x))\right]}.
	\label{xi}\ee

\medskip
\footnote{\it Now we would like, later, to exploit the fact that in the
unitary gauge the Higgs field has only $N$ components, while its
effective action is still of order $N^2$. In the large $N$ limit this
implies that the fluctuations $\varphi$ are suppressed, and $\phi$ is
in all respect a classical field. 

In particular it will be possible to apply the saddle point method on
its effective action:
	\be \e^{-\Gamma[\phi]} = \int\D\amu\e^{-{1\over\hbar}\left[S_{cl}
	+S_{quadr}-\hbar\delta(0)\int\di^3x\log\Delta^2(\phi(x))\right]}.\ee

Of course this calculation is still not simple, because it depends on the
classical field $\hat\A$.}

To this aim, we recall the action of fluctuations
	\be\S_{quadr} = N\tr\int \left[([\hat\Dmu,\anu]-[\hat\Dnu,\amu])^2
			+{g\over2}\hat G_{\mu\nu} [\amu,\anu] 
			+(g[\amu,\hat\phi+\varphi]+[\hat\Dmu,\varphi])^2+
				  \right.\0\\
			\left.+ \null^t\varphi V''
                                         (\hat\phi) \varphi \right]~~~~~~~
				\label{S2}\ee
understanding that $\varphi$ is diagonal.

\medskip
All the effect that we want to discuss arises from the massive gauge fields
circulating in a loop, so we calculate the divergent part of it.
	
Because the gauge propagator is constant at large momentum,
the loop contributes with an arbitrary number of insertions of interaction
terms.

The cubic interactions in the gauge or Higgs fields do not 
enter at one loop, and simultaneous interactions of two $\amu$ with a
Higgs field plus a background gauge give perturbative corrections to what
we need, so we leave them apart.

From the above action (\ref{S2}) we take the relevant
quadratic interaction terms of the fluctuating $\amu^{ij}$
with the external fields:
	\be g^2\amu^{ij}\amu^{ji}(2\hat\phi^\infty_i-2\hat\phi^\infty_j
			+\varphi_i'-\varphi_j')
			(\varphi_i'-\varphi_j')\ee
			
where the Higgs field $\phi=\hat\phi+\varphi$ has been decomposed
in a different way: the asymptotic constant field $\hat\phi^\infty$
which regulates the gauge bosons mass, plus the remaining classical
nonuniform background and fluctuating fields which have to be treated
as a total external field $\varphi'$: $\phi=\hat\phi^\infty+\varphi'$.

\FIGURE{\epsfig{file=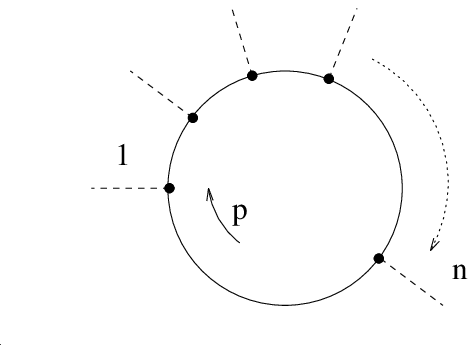}}
The loop consists thus of the same charged $\amu^{ij}$ field running along
the loop with propagator
	\be{-\gmunu\over p^2-m^2_{ij}}+
		{1\over m^2_{ij}}{p_\mu p_\nu\over p^2-m^2_{ij}}\ee
(with $m_{ij}=g|\hat\phi^\infty_i-\hat\phi^\infty_j|$)
and an arbitrary number of insertions of
$v_{ij}=g^2(2\hat\phi_i^\infty-2\hat\phi_j^\infty+\varphi_i'-\varphi_j')
		(\varphi_i'-\varphi_j')$.

We insert this $v_{ij}$ at zero external momentum, because we are dealing with
the divergent part. We have thus, for n insertions,
	\be \sum_{i<j} v_{ij}^n
	\int\di^3p\tr\left( {-\id\over p^2-m^2_{ij}}+
	{1\over m^2_{ij}}{\p\times\p\over p^2-m^2_{ij}}\right)^n=
	\delta(0)\sum_{i<j}\left({v_{ij}\over m_{ij}^2}\right)^n\ee

There is a combinatorial factor which comes from the $(n-1)!$ ways to insert
the interactions compared with the $n!$ ways to attach the resulting
counterterm. Hence the factor is $1/n$.

Summing up all these divergent contributions we reconstruct the logarithmic potential:
	\be &\hbar\delta(0)\sum_{i\neq j}\sum_n {1\over n}
	\left({g^2\left((2\hat\phi^\infty_i-2\hat\phi^\infty_j
			+\varphi_i'-\varphi_j')(\varphi_i'-\varphi_j')\right)
	\over m^2_{ij}}\right)^n=&\0\\
	&\hbar\delta(0)\sum_{i\neq j}\left(\log|\phi_i-\phi_j|
		-\log|\hat\phi^\infty_i-\hat\phi^\infty_j|\right)
		&\ee
\medskip
It clearly cancels only part of the Faddeev-Popov Vandermonde determinant
above in (\ref{xi}), and leaves the second term
function of the Higgs vacuum only:

	\be-\hbar\delta(0)\int\log\Delta^2(\hat\phi^\infty)\label{urp}\ee

Now, on one hand it is just a constant which is the same for all the
topological sectors and can be absorbed in the normalization for what
regards $\xi$ and all fluctuations, on the other it can be thought as
a potential correcting the vacuum constant value $\hat\phi^\infty$.

\medskip
This result needs some discussion. A similar mechanism is
well known from the study of the perturbative corrections to the
Higgs potential \cite{Sher}: the quantum corrections keep a
nonzero Higgs v.e.v. also in the limit of no bare breaking $\mu\to0$. 

From this point of view it provides a repulsive potential for the
eigenvalues of the $\phi^\infty$ v.e.v., justifying the assumption of 
$\hat\phi_i\neq\hat\phi_j$ for the vacuum value.

\medskip
It may appear strange that we have obtained an effective potential with a
divergent $\delta(0)\simeq\Lambda^3$ constant, because it seems to be
stronger than any of the renormalized other terms in the action, in the
continuum limit.

Nonetheless the fact that it depends only on $\hat\phi^\infty$ does not
allows us to treat it like the other terms in the effective action. 
The scalar potential $V(\phi)$ is of a very different nature because
it depends on the fluctuations also. It can lead the field to attain
its minimum, which has to be the vacuum for the system, in order
to be stable against local perturbations. 

The effective potential (\ref{urp}) is expected to play a role
let's say, just at the moment of symmetry breaking, when the sources
should decide which direction in the Cartan space to choose.
It is at this stage that the Vandermonde potential is important and the
quantum theory requires one unique direction with nondegenerate Higgs
field.

Eventually one must leave to external sources just the discrete
choice among the $N(N-1)$ possible vacuums related by Weyl.


\medskip
All this analysis is independent of the large $N$ limit, but for our 
dilute gas of monopoles, we must would like to know $\phi^\infty$. 
In the next paragraph we will find it according to the above
discussion, for large $N$.

\subsection{The Higgs vacuum}

The vacuum field $\phi^\infty$ plays an important role in the
dilute gas picture, because it decides if the monopoles are
relevant to confinement.

\medskip
According to the discussion about the unitary gauge in last section, the
vector of eigenvalues $\ph^\infty\in\R^{N-1}$ is defined by the minimum of
	\be -\sum_{i<j} \log|\phi^\infty_i-\phi^\infty_j|\label{rhoeq}\ee
with the constraint
	\be \sum_i (\ph^\infty_i)^2=\mu^2\label{constr}\ee
and we recall that $\mu^2$ is of order $N$, so that the components of
$\ph^\infty$ are of order 1. 

\medskip
The solution for finite $N$, although existing, is not illuminating.
We instead turn to the large $N$ limit and introduce the
(non standard) density of eigenvalues:
\be \rho(\phi^\infty) = \left(N {\di \phi^\infty_i\over\di i}\right)^{-1}\ee
It is of order $1$, as a consequence of last constraint. 

We want to solve for it to obtain the distribution of
eigenvalues of the vacuum Higgs field. The equation for
$\rho(x)$ is, from (\ref{rhoeq}), (\ref{constr}):

	\be N \cdot P\int_{-a}^a {\rho(x)\over x-y}\di y=2\lambda x \ee
where the right hand side is the Lagrange multiplier for the constraint
(\ref{constr}) which is equivalent to $\int x^2\rho(x)\di x=\mu$.

In the large $N$ limit the Lagrange
multiplier is negligible (we do not expect an attraction of
eigenvalues coming from this constraint, because $\mu^2\sim N$)
and we have to solve: 
	\be P\int_{-a}^a {\rho(x)\over x-y}\di y=0 \ee

The by now standard method \cite{BrItPaZub} is to introduce the
resolvent of $\rho$ as 
	\be F(x)=P\int_{-a}^a {\rho(y)\over x-y}\di y \ee
which has the following properties: It is analytic out of the {\it
cut} $[-a,a]$ on the real axis; it goes to zero at infinity as $1\over
|x|$; It is real on the real axis $[-a,a]$ excluded; near the cut it
has zero real part and a discontinuity in the imaginary part given by 
the unknown $\pi \rho(x)$. 

The unique function with these requirements is 
	\be F(x) = C {1\over \sqrt{x^2-a^2}}\ee
from which we finally read (and normalize) the distribution $\rho(x)$:
	\be \rho(x) = {1\over\pi} {1\over\sqrt{a^2-x^2}}.\label{rho}\ee
The result is thus an {\it inverted semicircle law}.

\medskip
Its domain is defined by the constraint (\ref{constr}) 
(we have introduced the fixed scale $\tilde\mu^2=\mu^2/N$):
	\be N \int_{-a}^a x^2 \rho(x) dx=N\tilde\mu^2\ee
which gives:
		\be a^2 = {2\over \pi} \tilde\mu^2\ee
\bigskip
$\rho(x)$ of (\ref{rho}) represents the following Higgs configuration
in the large $N$ limit:
	\be \phi^\infty_i = \sqrt{2\over\pi}\tilde\mu \cos(\pi{i\over N}).
			\label{VEV}\ee
Of course there are $N!$ equivalent configurations related by Weyl
permutations. We show the ordered ones in the picture.

\FIGURE{\epsfig{file=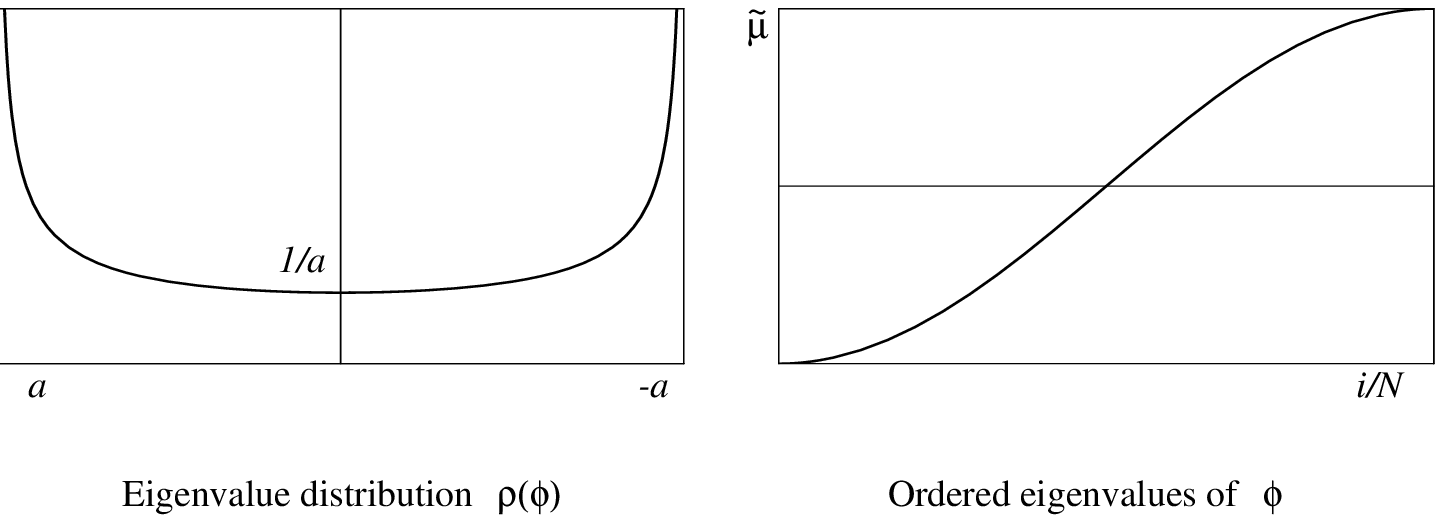}}
 
Let's make a few comments on this result:

$\bullet$ First, in the large $N$ limit we find that the Higgs
eigenvalues remain of order one, while in the usual picture
there is no indication and one could also assign
all the vacuum expectation value $\mu$ to a single eigenvalue.

$\bullet$ Second, the Weyl degeneracy is completely broken by
$\ph^\infty$, and no symmetry remains apart from the abelian $U(1)^{N-1}$.

$\bullet$ Third, and more important, our vacuum value (\ref{VEV})
shows a peculiar characteristic, namely that, near the edges of the
distribution, differences of eigenvalues are of order $1\over N^2$.
As a consequence the masses of gauge bosons in the $SU(2)$ sectors near
the two ends are vanishing as $1/N^2$.In the middle one has the normally expected
$1/N$ masses.

This facts means that there will be an infinite number
of very light monopoles, with masses vanishing as $1/N$.

The implications on the semiclassical picture of monopole gas are
interesting, and we will deal with them in the next section.

\medskip
On the other hand, independently of the monopole gas, the presence of an
infinity of massless modes can lead to a new phase of the model, of
course present only in the $N=\infty$ sector.

\bigskip
Let us mention also a nice correspondence with the solvable cases of
$\cN=2\to \cN=1$ supersymmetric gauge theories in four dimensions.

In the $SU(N)$ generalization of the $(\cN=1){SYM}_4$, as analyzed by
Douglas and Shenker \cite{DouShe}, the $\ph^\infty$ field is represented
by the points in the moduli space of the $\cN=2$ theory which become the
vacua of the theory broken to $\cN=1$. 

The exact solution shows that those points are exactly of the form of
our $\ph^\infty$. In particular the Weyl degeneracy is completely
broken, and moreover the different charged sectors will have
masses of gauge bosons and of monopoles given again by
$\cos(\pi{i\over N})-\cos(\pi{j\over N})$, for example $\simeq1/N^2$
near the edges of the distribution.

We remark that in the supersymmetric theory the ground state appears in
a full nonperturbative and geometric way, while in our case it must be
stressed that we are not dealing exactly with quantum corrections, but
with a global factor in the Higgs measure, expected to be important just
during the formation of the system.

A main feature, the hierarchy of masses, is the same, and this is at
least curious.

\subsection{The hierarchy of masses and monopole degeneracies}

Here we come to the observation that in the large $N$ limit there is
a hierarchy of masses which arises from the model, once given the
distribution of Higgs eigenvalues. As $N$ gets large, many of the
objects become very light, potentially bringing in a different physics, like
happens in other cases in this limit.

\medskip
We have the gauge bosons and monopole pseudo-masses
	\be m_W(\q) = g|\ph^\infty\cdot\q|\0\qquad 
	    \S_{cl}(\q) = 4\pi N {|\ph^\infty\cdot\q|\over g}\0\ee
which are function of the $SU(2)$ root $\q$ which specifies the color
sector. Explicitly as a function of the Higgs eigenvalues
	\be|\ph^\infty\cdot\q|=g{1\over\sqrt2}|\ph^\infty_i-\ph^\infty_j|\ee
when $\q$ is the charge in the sector $(ij)$.

The possible values are $N(N-1)$ and we see that according to the
distribution of eigenvalues in the Higgs field, they range in
the interval $(\mu/N^2,1)$.

Moreover at the ends of the Higgs domain, $\pm\mu$, where the eigenvalues
concentrate, we have masses of order $1/N^2$, while from
the standard differences we get $\sim N^2$ masses of order $1/N$ and more.

We can look at the distribution of differences:
	  \be \sigma(d)&=&\int\int dxdy\rho(x)\rho(y)\delta(|x-y|-d)\0\\
		    &=&4\int_{-a}^{-d/2}dt{1\over
			\sqrt{(a^2-t^2)(a^2-(t+d)^2)}}.\ee
which is logarithmic singular for $d\to0$ showing the phenomenon:
		\be \sigma(d)\simeq {2\over \pi^2} {1\over a}\log a/d.\ee
because masses are proportional to $d$ this is also the distribution 
of masses in the model near the edges. 

\medskip
The monopole pseudo-mass $\S_{cl}$ has a $N$ factor which compensates this
vanishing in the middle part of the distribution, but still many
monopoles tend to have zero action in the limit.

\subsection{Dilute gas in the large $N$ limit}

We will use these informations in the dilute gas ensemble of section
{\bf\ref{gas}}. 

We have to perform the average of $\xi(\q)$ in the subspace of the
simple roots which are not orthogonal to a given weight
$\m_\al$. These are of the form: $\q=\m_\al\pm\m_\beta$ ($\beta\neq\al$).

Then
	\be \bar \xi(m_\al)={1\over (N-1)} \sum_{\beta\neq\al}
	\xi(\m_\al-\m_\beta)\ee

For large $N$ the average $\bar\xi$ has the form:

	\be \bar\xi\sim N^{9/2}\int dx\rho(x) {m_W^{9/2}\over g^3}
	A\e^{-N{m_W\over g^2}}\ee

which leads to the following average

	\be N^{9/2}A\int_{-a}^a dx{1\over\sqrt{a^2-x^2} }
		{((x-b)\tilde\mu)^{9/2}\over g^3}
	\e^{-N(x-b){\tilde\mu\over g^2}}\ee
where the large $N$ limit has to be performed. 

Here the first observation is that the exponential in the large $N$
limit becomes a delta function (times a $1/N$ factor).

But then the $(x-b)^{9/2}$ prefactor drives the integral to zero. In
fact the integral is expanded and factors of $1/N$ appear leaving derivatives
of the delta function. 

However we know that also the factor $A$ depends on $N$, being
currently an open problem. 

In case $A$ will turn out not to grow faster than $N^2$, the string
tension will be suppressed and vanish in the large $N$ limit. This
could be explained noticing that also in the large $N$ limit, as in
the BPS case, all the Higgs fields become massless. This important peculiarity
distinguishes between the large $N$ $YMH$ and {\it pure YM} theory at
$\lm>0$, whereas for $SU(2)$ the Higgs field just provides an explicit
scale for the monopole stability. A complete discussion will be
available by the time the nontrivial quantity $A$ in the one monopole 
determinant will be studied.

\bigskip

\appendix 
\section{A remark on the Gribov problem}

Here we want to make some remarks on the Gribov ambiguity in the
present case where the unitary gauge is chosen.

It turns out that no Gribov problem is present, and the unitary gauge
is thus just a feasible gauge with some unusual poperties.
Moreover the absence of unphysical fields, and the fact that the Higgs field
is already diagonal, makes it a good tool to investigate the quantum theories.

\medskip
The main point observation in the Gribov problem 
is that the Faddeev-Popov determinant carries the
information on how good is our choice of gauge.
It represents the local jacobian for the change of variables from the
connection space to the quotient by the gauge group action where we
have our physical theory defined.

At points in functional space where the FP determinant vanishes,
it means that we are not taking a complete quotient, in fiber bundle
language the section we have chosen becomes tangent to the fiber
along some gauge direction.

These points usually mark what is called the Gribov horizon, the name 
because it is the boundary of the maximal region where a section
can be continued.

\medskip
Here, to be concrete, we have in three dimensions a gauge connection
and a matter field so the space of fields $\Gamma$ is that of couples
$(A_\mu,\phi)$, functions from $\R^3$ to the gauge algebra.
As we have seen $\Gamma$ is disconnected in components according to
the total magnetic charge, as described in section {\bf \ref{ym3}}.

In every component acts the gauge group $\cg=\{g(x):\R_3\to SU(N)\}$,
and no boundary conditions have to be imposed at infinity because by
homotopic arguments gauge transformations connected to the identity
do not change component.

This is best and more appropriately seen in the regular gauges, while in
the unitary gauge we know that $A_\mu$ becomes singular along some Dirac
string (still with $\S_{cl}<\infty$). Nevertheless the unitary gauge
has nice properties, namely the FP determinant and the gauge fixing 
only depend on the Higgs field, so that one can draw some conclusion.

\medskip
The unitary gauge introduced in section~\ref{qu}, has a 
Faddeev-Popov determinant which turns out to be
	\be \prod_x \prod_{ij}\left(\phi_i(x)-\phi_j(x)\right)^2.\ee

Now we recall, as remarked at the end of in section~\ref{ym3}, 
that in nontrivial sectors of the Higgs winding at infinity, 
$\phi$ has necessarily some coinciding eigenvalues at some point.

This is proven in any regular gauge but is valid also in other gauges
because it's a gauge invariant statement.

In the semiclassical picture of monopole gas, at each monopole location
two (or more) eigenvalues coincide.

Hence we find that in the unitary gauge the FP determinant above
seems to vanish identically for any nontrivial configuration.
More suggestively one can think that the Gribov horizon is made
of monopole configurations. 
The same happens at each field configuration throughout the whole
nontrivial sectors.

This would mean that the unitary gauge is an ill defined section,
in that it does not fix the gauge. Explicitly it would leaves
intact some gauge subgroup at the points in space where we have a monopole.

\medskip
However there are two points which solve this seemingly bad problem.

$\bullet$ The determinant in the form above is ill defined. It requires
us to live in a distribution space, whereas we usually consider smooth
functions.
After this remark it appears evident that no gauge invariance
remains unfixed, because the smoothness constrains the gauge variation
at the "origin" to follow that in the neighbor. So there is no such
thing as the gauge variation at a single point, even if the theory has
local invariance.

$\bullet$ In subsection~\ref{qu} we proved that the Faddeev-Popov determinant
and its vanishing is canceled by the gauge loops, so that if the jacobian
vanishes it is just because the change of variables is singular, and in
fact at the same time the integral takes care of this and diverges by the
same amount so to correct the measure.

This is different from the standard Gribov phenomenon where the vanishing of
the FP determinant is a unavoidable problem in the context of a
renormalizable theory.

\newpage

\bibliographystyle{unsrt} \bibliography{bib}

\begin{thebibliography}{10}

\bibitem{GeoGla}
H.~Georgi and S.~Glashow.
\newblock {\em Phys. Rev. Lett.}, 28:1494, 1972.

\bibitem{Hoo74}
G.~`tHooft.
\newblock {\em Nucl. Phys.}, B79:2766, 1974.

\bibitem{Poly74}
A.M. Polyakov.
\newblock {\em JEPT Lett.}, 20:894, 1974.

\bibitem{Dirac}
P.A.M. Dirac.
\newblock {\em Physical Review}, 74:817, 1948.

\bibitem{tHooft}
G.~'t~Hooft.
\newblock Topology of the gauge condition and new confinement phases in
  non-abelian gauge theories.
\newblock {\em Nuclear Physics}, B190:455--478, 1981.

\bibitem{NieOle}
H.B. Nielsen and P.~Olesen.
\newblock {\em Nucl. Phys.}, B61:45, 1973.

\bibitem{Weinberg}
S.~Weinberg.
\newblock {\em Progr. of Teor. Phys. Suppl.}, 86:43, 1986.

\bibitem{conftHo}
G.~`t~Hooft.
\newblock in ''high energy physics``, eps international conference.
\newblock Palermo 1975, ed. A. Zichichi.

\bibitem{confMan}
S.~Mandelstam.
\newblock {\em Phys. Rep.}, 23C:245, 1976.

\bibitem{confPar}
G.~Parisi.
\newblock {\em Phys. Rev.}, D11:971, 1975.

\bibitem{DGL-conf}
H.~Toki, H.~Suganuma, S.~Sasaki, and H.~Ichie.
\newblock Dual ginzburg-landau theory for quark confinement and dynamical
  chiral-symmetry breaking.
\newblock (hep-ph/9502277).

\bibitem{DGL-conf2}
H.~Toki, H.~Suganuma, S.~Sasaki, and H.~Ichie.
\newblock Dual ginzburg-landau theory for nonperturbative qcd.
\newblock (hep-th/9506366).

\bibitem{Poly75}
A.M. Polyakov.
\newblock Compact gauge fields and the infrared catastrophe.
\newblock {\em Physics Letters}, 59B(1):82, 1975.

\bibitem{Poly77}
A.M. Polyakov.
\newblock Quark confinement and topology of gauge theories.
\newblock {\em Nuclear Physics}, B120:429--458, 1977.

\bibitem{AtiyahHitchin}
M.~Atiyah and N.~Hitchin.
\newblock {\em The geometry and dynamics of magnetic monopoles}.
\newblock Princeton University Press, 1988.

\bibitem{digiacMCTV}
Adriano Di~Giacomo.
\newblock Monopole condensation in gauge theory vacuum.
\newblock (Talk given at Lattice 95?).

\bibitem{DasWadia94}
Sumit~R. Das and Spenta~R. Wadia.
\newblock Quark confinement in 2+1 dimensional pure ym theory.
\newblock (hep-th/9503184).

\bibitem{largeN1}
G.~'t~Hooft.
\newblock {\em Nucl. Phys.}, B72:461, 1972.

\bibitem{largeN2}
L.G. Jaffe.
\newblock {\em Rev. Mod. Phys.}, 54:407, 1982.

\bibitem{kazwin}
V.A. Kazakov and T.~Winter.
\newblock Large $n$ phase transition in the heat kernel on the $u(n)$ group.
\newblock (LPTENS-94/28), 1994.

\bibitem{DasWadia}
Spenta~R. Wadia and Sumit~R. Das.
\newblock Topology of quantum gauge fields and duality (i): Yang-mills-higgs
  system in 2+1 dimensions.
\newblock {\em Physics Letters}, 106B:386, 1981.

\bibitem{DouShe}
M.R. Douglas and S.H. Shenker.
\newblock Dynamics of $su(n)$ supersymmetric gauge theory.
\newblock (hep-lat/9503163).

\bibitem{Oleszcz}
M.~Oleszczuk.
\newblock Mass generation in three dimensions.
\newblock (hep-th/9412049).

\bibitem{det1}
C.~Callias and C.H. Taubes.
\newblock Functional determinants in euclidean yang-mills theory.
\newblock {\em Comm. Math Phys.}, 77:229, 1980.

\bibitem{det2}
D.I. D'yakonov, V.Yu. Petrov, and A.V. Yung.
\newblock Quasiclassical expansion in external yang mills fields and the
  approximate calculation of functional determinants.
\newblock {\em Sov. J. Nucl. Phys.}, 39:150, 1984.

\bibitem{zarembo}
K.~Zarembo.
\newblock Monopole determinant in ym theory at finite temperature.
\newblock (hep-th/9510031).

\bibitem{HooVel}
G.~\'tHooft and M.~Veltman.
\newblock {\em Nucl. Phys.}, B44:189, 1972.

\bibitem{Li}
Ling-Fong Li.
\newblock Group theory of the spontaneously broken gauge symmetries.
\newblock {\em Physical Review D}, 9(6):1723, 1974.

\bibitem{Sch}
A.~Schwarz.
\newblock {\em Topology for Physicists}.
\newblock Springer Verlag, 1993.

\bibitem{Arafune}
J.~Arafune, P.G.O. Freund, and C.J. Goebel.
\newblock Topology of higgs fields.
\newblock {\em Journal of Mathematical Physics}, 16(2):433--437, February 1975.

\bibitem{goddolive}
P.~Goddard and D.~Olive.
\newblock Magnetic monopoles in gauge field theories.
\newblock {\em Rep. Prog. Phys.}, 41:1357, 1978.

\bibitem{WilkGold}
David Wilkinson and Alfred~S. Goldhaber.
\newblock Spherically symmetric monopoles.
\newblock {\em Physical Review D}, 16(4):1221, 1977.

\bibitem{monSU4}
Y.~Brihaye and J.~Nuts.
\newblock Magnetic monopoles in $su(4)$ gauge theories.
\newblock {\em Journal of Mathematical Physics}, 18(11):2177, 1977.

\bibitem{monSU3sinha}
A.~Sinha.
\newblock Su(3) magnetic monopoles.
\newblock {\em Physical Review D}, 14(8):2016, 1976.

\bibitem{monSU3corri}
E.~Corrigan, D.I. Olive, D.B. Fairlie, and J.~Nuyts.
\newblock Magnetic monopoles in $su(3)$ gauge theories.
\newblock {\em Nuclear Physics}, B106:475, 1976.

\bibitem{WuYang}
Tai~Tsun Wu and Chen~Ning Yang.
\newblock Concept of nonintegrable phase factors and global formulation of
  gauge fields.
\newblock {\em Physical Review D}, 12(12):3845, 1975.

\bibitem{EnglWind}
F.~Englert and P.~Windley.
\newblock Quantization condition for 't hooft monopoles in compact simple lie
  groups.
\newblock {\em Physical Review D}, 14(10):2728, 1976.

\bibitem{BPS}
M.K. Prasad and C.M. Sommerfeld.
\newblock Exact classical solution for the 't hooft monopole ad the julia-zee
  dyon.
\newblock {\em Phys. Rev. Lett.}, 35(12):760, 1975.

\bibitem{TetraCubic}
Conor~J. Houghton and Paul~M. Sutcliffe.
\newblock Tetrahedral and cubic monopoles.
\newblock (hep-th/9601146).

\bibitem{JaffeTau}
Arthur Jaffe and Clifford Taubes.
\newblock {\em Vortices and Monopoles}.
\newblock Birkhauser, 1980.

\bibitem{PolyBook}
A.M. Polyakov.
\newblock {\em Gauge fields and strings}.
\newblock Harwood Academic Publisher, 1987.

\bibitem{Sher}
M.~Sher.
\newblock Electroweak higgs potentials and vacuum stability.
\newblock {\em Physics Reports}, 179:273, 1989.

\bibitem{Hoo1}
G.~`tHooft.
\newblock {\em Nucl. Phys.}, B33:823, 1971.

\bibitem{Hoo2}
G.~`tHooft.
\newblock {\em Nucl. Phys.}, B35:167, 1971.

\bibitem{Lee}
B.W. Lee.
\newblock {\em Phys. Rev. D}, 5:823, 1972.

\bibitem{LeeZin}
B.W. Lee and J.~Zinn-Justin.
\newblock {\em Phys. Rev. D}, 5:3121,3137,3155, 1972.

\bibitem{LeeYan}
T.D. Lee and C.N. Yang.
\newblock {\em Phys. Rev.}, 128:885, 1962.

\bibitem{unitgaugeQED}
Giuseppe Bimonte, Roberto Iengo, and G.~Lozano.
\newblock Uv asymptotycally free qed as broken ym theory in the unitary gauge.
\newblock (hep-th/9502185).

\bibitem{AppQui}
T.~Appelqyust and H.~Quinn.
\newblock Divergence cancellations in a simplified weak interaction model.
\newblock {\em Phys. Lett.}, 39B(2):229, 1972.

\bibitem{BrItPaZub}
E.~Brezin, C.~Itzykson, G.~Parisi, and J.B. Zuber.
\newblock Planar diagrams.
\newblock {\em Commun. Math. Phys.}, 59:35--51, 1978.

\bibitem{EK}
T.~Eguchi and H.~Kawai.
\newblock {\em Phys. Rev. Lett.}, 48:1063, 1982.

\bibitem{Parisi}
G.~Parisi.
\newblock {\em Phys. Lett.}, 112B:463, 1982.

\bibitem{BhaHelNeu}
G.~Bahnot, U.~Heller, and H.~Neuberger.
\newblock The quenched eguchi-kawai model.
\newblock {\em Phys. Lett.}, 113B:47--50, 1982.

\bibitem{GroKit}
D.~Gross and Y.~Kitazawa.
\newblock A quenched momentum prescription for large-$n$ theories.
\newblock {\em Nuclear Phys.}, B206:440--472, 1982.

\bibitem{Das}
S.M. Das.
\newblock Some aspects of large-$n$ theories.
\newblock {\em Review of Modern Physics}, 59(1):235--261, 1987.

\bibitem{Bars}
I.~Bars, M.~Gunaydin, and S.~Yankielowicz.
\newblock Reduced chiral $u(n)\times u(n)$ model.
\newblock {\em Nucl. Phys.}, B219:81--115, 1983.

\bibitem{douglas}
Michael~R. Douglas.
\newblock Conformal field theory techniques for large $n$ group theory.
\newblock (hep-th/9303159).

\bibitem{menotti}
P.~Menotti and E.~Onofri.
\newblock The action of $su(n)$ lattice gauge theory in terms of the heat
  kernel on the group manifold.
\newblock {\em Nucl. Phys.}, B190:288--300, 1981.

\bibitem{minpol}
J.A. Minahan and A.P. Polychronakos.
\newblock Classical solutions for two dimensional qcd on the sphere.
\newblock (hep-th/9309119).

\bibitem{maty}
A.~Matytsin.
\newblock On the large $n$ limit of the itzykson-zuber integral.
\newblock (PUPT-1405), 1996.

\bibitem{GroWit}
D.~Gross and E.~Witten.
\newblock Possible third order phase transition in the large $n$ lattice gauge
  theory.
\newblock {\em Phys. Rev.}, 21(2):446, 1980.

\end{thebibliography}

\end{document}